\documentclass[aps, prd, amsmath, floats, floatfix, twocolumn, preprintnumbers, nofootinbib, showpacs, eqsecnum]{revtex4-1}
\usepackage{graphicx}
\usepackage{color}

\newcommand{\be}{\begin{eqnarray}}
\newcommand{\ee}{\end{eqnarray}}
\newcommand{\rar}{\rightarrow}

\begin{document}

\title{Towards the use of the most massive black hole candidates in AGN\\ 
to test the Kerr paradigm}

\author{Cosimo Bambi}
\email{Cosimo.Bambi@physik.uni-muenchen.de}
\affiliation{Arnold Sommerfeld Center for Theoretical Physics\\
Ludwig-Maximilians-Universit\"at M\"unchen, 80333 Munich, Germany}

\date{\today}

\begin{abstract}
The super-massive objects in galactic nuclei are thought to be the Kerr black holes 
predicted by General Relativity, although a definite proof of their actual nature is still 
lacking. The most massive objects in AGN ($M \sim 10^9$~$M_\odot$) seem to 
have a high radiative efficiency ($\eta \sim 0.4$) and a moderate mass accretion 
rate ($L_{\rm bol}/L_{\rm Edd} \sim 0.3$). The high radiative efficiency could suggest 
they are very rapidly-rotating black holes. The moderate luminosity could indicate 
that their accretion disk is geometrically thin. If so, these objects could be excellent 
candidates to test the Kerr black hole hypothesis. An accurate measurement of the 
radiative efficiency of an individual AGN may probe the geometry of the space-time 
around the black hole candidate with a precision comparable to the one achievable 
with future space-based gravitational-wave detectors like LISA. A robust evidence 
of the existence of a black hole candidate with $\eta > 0.32$ and accreting from 
a thin disk may be interpreted as an indication of new physics. For the time being, 
there are several issues to address before using AGN to test the Kerr paradigm, 
but the approach seems to be promising and capable of providing interesting results 
before the advent of gravitational wave astronomy. 
\end{abstract}


\maketitle


\section{Introduction}

Gravity has been tested and verified for distances in the range $\sim$1~mm to 
$\sim$1~pc (mainly within its Newtonian limit) and for weak gravitational 
fields~\cite{w,5th}. The research is now moving to check the validity of the 
theory at cosmological scales, sub-millimeter distances, and for strong gravitational 
fields. One of the most intriguing predictions of General Relativity (GR) is 
that the collapsing matter produces singularities in the space-time. According 
to the weak cosmic censorship conjecture, singularities of gravitational collapse 
must be hidden within black holes (BHs)~\cite{p}. In 4-dimensional GR, 
uncharged BHs are described by the Kerr solution, which is completely 
specified by two parameters, the mass, $M$, and the spin angular momentum, 
$J$~\citep{c}. The condition for the existence of the event horizon is $a_* 
\le 1$, where $a_* = |J/M^2|$ is the spin parameter\footnote{Throughout the 
paper I use units in which $G_N = c = 1$, unless stated otherwise.}. When 
$a_* > 1$, there is no horizon and the central singularity is naked, violating 
the weak cosmic censorship conjecture.

Astronomers have discovered at least two classes of BH candidates (for a review,
see e.g. Ref.~\cite{n}): stellar-mass objects in X-ray binary systems ($M \sim 5 - 
20$~$M_\odot$) and super-massive objects in galactic nuclei ($M \sim 10^5 - 
10^9$~$M_\odot$). The estimates of the masses of these objects are robust, because 
determined via dynamical measurements and without any assumption about the 
geometry of the space-time. The key-point is that the stellar-mass objects in X-ray 
binary systems are too heavy to be neutron or quark stars for any reasonable matter 
equation of state~\cite{kb}, while the super-massive objects at the 
centers of galaxies are too heavy, compact, and old to be clusters of non-luminous 
bodies, as the cluster lifetime would be shorter than the age of these systems~\cite{m}. 
All these objects are therefore thought to be the BHs predicted by GR, as 
they cannot be explained otherwise without introducing new physics. There are also 
some observations interpreted as an indirect evidence for the existence of the event 
horizon~\cite{bln} (but see~\cite{a}). On the contrary, there is no indication that 
the geometry around these objects is described by the Kerr metric.

Testing the Kerr BH hypothesis is thus the next step to progress in this research
field and several authors have indeed suggested possible ways to do it using 
present and future data (for a review, see e.g. Ref.~\cite{b4}). A very promising approach 
is the detection of extreme mass ratio inspirals (EMRIs, i.e. systems consisting 
of a stellar-mass compact object orbiting a super-massive BH candidate) with 
future space-based gravitational-wave antennas. Missions like LISA will be able 
to follow the stellar-mass compact object for millions of orbits around the central 
super-massive BH candidate, and therefore deviations from the Kerr geometry 
will lead to a phase difference in the gravitational waveforms that grows with the 
number of observed cycles~\cite{r}. However, these data will not be 
available shortly, as the first mission will be at best in the early 2020s. 
The nature of BH candidates can also be tested by extending the methods currently 
used to estimate the spin of these objects, such as X-ray continuum~\cite{bb1} 
and K$\alpha$-iron measurements~\cite{pj}, observations of quasi-periodic 
oscillations~\cite{jpp}, and measurements of the cosmic X-ray background~\cite{b2,b5}. 
These methods can in principle be applied even with present data, provided 
that the systematic errors are properly understood. Future observations of the 
shadow of nearby super-massive BH candidates are another exciting possibility 
to test the Kerr BH paradigm~\cite{bf}.

Previous studies have clearly pointed out that ``rapidly-rotating'' objects are the
best candidates to test the Kerr BH hypothesis: if the object rotates fast, even a
small deviation from the Kerr background can cause significant differences in the
properties of the electromagnetic radiation emitted by the gas of the accretion 
disk and peculiar features, otherwise absent in the Kerr geometry, may show 
up~\cite{bb2,bm}. The aim of this paper is to investigate the potentialities 
of the most massive BH candidates in AGN, through the measurement of their
radiative efficiency $\eta$, defined by $L_{\rm bol} = \eta \dot{M}$, where
$L_{\rm bol}$ is the bolometric luminosity of the source and $\dot{M}$ is the mass
accretion rate of the BH candidate. The estimate of the mean radiative efficiency
of AGN through the Soltan's argument~\cite{s} already suggests the presence
of rapidly-rotating BHs~\cite{erz,ho}. Recently, Davis and Laor have proposed a way to
measure the radiative efficiency of individual AGN~\cite{dl}. The authors found that the
most massive BH candidates (with a mass $M \sim 10^9$~$M_\odot$) would
have a high radiative efficiency, up to $\eta \sim 0.4$, and a moderate mass 
accretion rate, $L_{\rm bol}/L_{\rm Edd} \sim 0.3$, where $L_{\rm Edd}$ is the
Eddington luminosity of the source. The standard accretion disk model in a Kerr
background would predict a high value of the spin parameter $a_*$ for these 
objects, extremely close to 1. At the same time, the moderate luminosity (in 
Eddington units) may indicate a thin accretion disk and the applicability of the 
standard accretion disk model. If these estimates and these considerations
are correct, the most massive compact objects in AGN would be excellent 
candidates to test the Kerr paradigm. The sole measurement of $\eta$ can
potentially constrain either $a_*$ and a deviation from the Kerr geometry.

The paper is organized as follows. In Section~\ref{s-2}, I review the standard
accretion disk model, its assumptions and properties, its effects on the 
evolution of the spin parameter of the central object, and its applicability. 
In Section~\ref{s-k}, I consider an accretion disk in the Kerr background, 
summarizing well-known results that should be expected if the BH candidates 
are the BHs of GR. In Sections~\ref{s-jp} and \ref{s-mn}, I discuss accretion 
disks respectively in the Johannsen-Psaltis (JP)~\cite{jp} and in the Manko-Novikov 
(MN)~\cite{mn} space-times. These are two metrics that can be conveniently used to 
describe a background deviating from the Kerr geometry. The nature of the 
two metrics is definitively different: the JP metric describes non-Kerr BHs in 
a putative alternative theory of gravity, while the MN one is an exact solution 
of the Einstein's vacuum equations and can describe the exterior gravitational 
field of generic compact objects. In both cases, the body is characterized by 
a mass, a spin angular momentum, and an infinite number of  ``deformation 
parameters'', even if here, for the sake of simplicity, I will consider only a single 
deformation parameter at a time. I will show that the two metrics present 
common features. In particular, a high radiative efficiency necessarily requires 
a very rigid compact object, much stiffer than a self-gravitating fluid with ``normal'' 
equations of state. The confirmation of the existence of individual AGN with 
high radiative efficiency ($\eta > 0.3$) can potentially either be used to put 
strong constraints on the Kerr nature of astrophysical BH candidates and
to discover new physics, as in the framework of the standard accretion disk 
model $\eta$ cannot exceed 0.32. In Section~\ref{s-d}, I discuss the findings
of this paper in relation with current estimates of the radiative efficiency of AGN. 
The conclusions and the issues that need to be addressed before using the 
most massive objects in AGN to really test the Kerr BH hypothesis are reported
in Section~\ref{s-c}. In Appendices~\ref{a-1} and \ref{a-2}, the reader can
find the non-vanishing metric coefficients respectively of the JP and of the
MN metric.

\section{Accretion disks}\label{s-2}

\subsection{Novikov-Thorne model}

The Novikov-Thorne (NT) model is the standard model for accretion disks~\cite{nt}. 
It describes geometrically thin and optically thick disks and it
is the relativistic generalization of the Shakura-Sunyaev (SS) model~\cite{ss}.
The disk is thin in the sense that the disk opening angle is $h = H/r \ll 1$,
where $H$ is the thickness of the disk at the radius $r$. Magnetic fields are
ignored. In the Kerr background, there are four parameters (BH mass $M$,
BH spin parameter $a_*$, mass accretion rate $\dot{M}$, and viscosity
parameter $\alpha$), but the model can be easily extended to any 
(quasi-)stationary, axisymmetric, and asymptotically flat space-time.
Accretion is possible because viscous magnetic/turbulent stresses and
radiation transport energy and angular momentum outwards. The model
assumes that the disk is on the equatorial plane and that the disk's gas 
moves on nearly geodesic circular orbits. For long-term accretions, the disk 
is adjusted on the equatorial plane as a result of the Bardeen-Petterson 
effect~\cite{bp}. That remains true even in non-Kerr backgrounds~\cite{b1}.
The assumption of nearly geodesic circular orbits requires that the radial 
pressure is negligible compared to the gravitational force of the BH. Heat 
advection is ignored (it scales as $\sim h^2$) and energy is radiated from 
the disk surface.

The key-ingredient of the NT model is that the inner edge of the disk is
at the innermost stable circular orbit (ISCO), where viscous stresses are
assumed to vanish. When the gas's particles reach the ISCO, they quickly
plunge into the BH, without emitting additional radiation. At first approximation,
the {\it total efficiency} of the accretion process is
\be\label{eq-eta}
\eta_{\rm tot} &=& 1 - E_{\rm ISCO} \, ,
\ee
where $E_{\rm ISCO}$ is the specific energy of the gas at the ISCO radius 
and depends uniquely on the background geometry. In general, the total
power of the accretion process is converted into radiation and kinetic 
energy of jet/wind outflows, so we can write $\eta_{\rm tot} = \eta + \eta_{\rm k}$.
$\eta$ is the {\it radiative efficiency} and can be inferred from the
bolometric luminosity $L_{\rm bol}$ if the mass accretion rate is known:
$L_{\rm bol} = \eta \dot{M}$. In this paper, I will assume $\eta_{\rm k} = 0$,
i.e. no gravitational energy of the gas is converted to kinetic energy of 
outflows. This is a conservative assumption for what follows.

As a consequence of the accretion process, the BH spin parameter evolves. 
Since the gas particles arriving at the ISCO plunge quickly onto the central 
object, without emission of additional radiation, the BH changes its mass 
by $\delta M = E_{\rm ISCO}\delta m$ and its spin angular momentum by 
$\delta J = L_{\rm ISCO} \delta m$, where $L_{\rm ISCO}$ is the specific 
angular momentum of the gas at the ISCO, while $\delta m$ is the gas rest-mass. 
The evolution of the spin parameter of the BH turns out to be governed by 
the following equation:
\be\label{eq-a}
\frac{da_*}{d\ln M} &=& \frac{1}{M} 
\frac{L_{\rm ISCO}}{E_{\rm ISCO}} - 2 a_* \, .
\ee
If the right hand side of Eq.~(\ref{eq-a}) is positive, the accretion process
spins the BH up. If it is negative, the BH is spun down. The equilibrium spin 
parameter $a_*^{\rm eq}$ is reached when the right hand side of Eq.~(\ref{eq-a}) 
vanishes and its value depends on the geometry of the space-time. As
discussed in~\cite{b5} and explained briefly also in Sec.~\ref{s-jp}, 
the value of $a_*^{\rm eq}$ 
we can infer from Eq.~(\ref{eq-a}) assuming that all the accreting matter 
has its angular momentum in the same direction provides the maximum
value for the spin parameter of the super-massive BH candidates in galactic
nuclei. This fact will be used to get the constraint $a_* < a_*^{\rm eq}$.

Actually, not all the radiation emitted by the disk can escape to infinity. A part of 
it leaves the disk, but it is then captured by the BH. Including this effect, 
Eqs.~(\ref{eq-eta}) and (\ref{eq-a}) become~\cite{t}
\be
\label{eq-eta2}
\eta &=& 1 - E_{\rm ISCO} - \zeta_{\rm E}\, , \\
\frac{da_*}{d\ln M} &=& \frac{1}{M} 
\frac{L_{\rm ISCO} + \zeta_{\rm L}}{E_{\rm ISCO} + \zeta_{\rm E}} - 2 a_* \, ,
\label{eq-a2}
\ee
where $\zeta_{\rm E}$ and $\zeta_{\rm L}$ take into account the radiation
captured by the BH and their expression can be found in~\cite{t}. There is also
a part of radiation that is emitted by the disk and returns to the disk as a consequence
of light deflection (returning radiation), interacting with the disk's particles and 
changing the flux profile emitted by the disk (see e.g. Ref.~\cite{li}).

\subsection{Validity of the model} \label{ss-2}

If we relax the assumption of vanishing stresses at the inner edge of
the disk, the radiative efficiency increases. The bolometric luminosity of the disk
becomes
\be
L_{\rm bol} &=& \eta \dot{M} = \nonumber\\
&=& g_{\rm ISCO} \Omega_{\rm ISCO}
+ (1 - E_{\rm ISCO} - \zeta_{\rm E}) \dot{M} \, ,
\ee
where $g_{\rm ISCO}$ 
($g_{\rm ISCO} \ge 0$) and $\Omega_{\rm ISCO}$ are respectively the torque 
and the angular velocity of the gas at the ISCO. Another crucial point of the
NT model is the $\alpha$ viscosity parameter: magnetic fields, which are thought
to drive turbulence in disks, may not behave like a local scalar viscosity.
If magnetohydrodynamics torques are present, the radiative efficiency $\eta$ 
can even exceed 1, as the flow taps the spin energy of the BH~\cite{ak}. 
In this case, it is not easy to recover $E_{\rm ISCO}$ from the estimate of $\eta$ 
and even the equilibrium spin parameter $a_*^{\rm eq}$ is not given by 
Eq.~(\ref{eq-a}), but it turns out to be lower~\cite{g}. GRMHD simulations show 
also that the gas's particles may not follow the geodesics of the space-time 
inside the ISCO~\cite{k} and that a significant emission of radiation from the 
plunging region is possible~\cite{n09}.

So, how good is the NT model to describe the accretion disk around astrophysical
BH candidates? For non-magnetized and weakly-magnetized disks, there 
is a common consensus that the NT model describes correctly thin disks, $h \ll 1$,
when the viscosity parameter is small, $\alpha \ll 1$~\cite{ap}. A common criterion 
to select sources with thin disks is that $L_{\rm bol}/L_{\rm Edd}$ does not exceed
0.3~\cite{mcc}. In the case of magnetized disks (in Kerr background, as there 
are no simulations for other space-times), the issue is still open and controversial.
The GRMHD simulations in Ref.~\cite{p10} (see also~\cite{p11}) show that the stress 
at the inner edge of the disk scales as $h$ and the authors conclude that the NT model 
with vanishing stress boundary condition is recovered for $h \rar 0$. The GRMHD
simulations in~\cite{n09} show instead large stress at the inner edge 
of the disk even when $h \rar 0$; according to these authors, the NT model cannot describe
magnetized disks, even when the disk is very thin. It is not clear if the disagreement
between the two groups can be attributed to different configurations of the
magnetic fields, different resolution of the simulations, or something else.

\begin{table*}
 \centering
\begin{tabular}{ccccccccccccc}
\hline
\hspace{0.5cm} &  $a_*$ & \hspace{0.5cm} & $E_{\rm ISCO}$ & \hspace{0.5cm} & $L_{\rm ISCO}$ & \hspace{0.5cm} & $\eta$ & \hspace{0.5cm} & $\Omega_{\rm ISCO}M$ & \hspace{0.5cm} & $r_{\rm ISCO}/M$ & \hspace{0.5cm} \\
\hline
& 0 && $\sqrt{8/9} $ && $\sqrt{12}$ && $1 - \sqrt{8/9} \approx 0.057$ && $1/\sqrt{6^3} \approx 0.068$ && 6 &\\
& 1 && $1/\sqrt{3}$ && $2/\sqrt{3}$ && $1 - 1/\sqrt{3} \approx 0.423$ && $1/2$ && 1 &\\ \\
& 0.99614 && 0.700 && 1.452 && 0.300 && 0.402 && 1.307 &\\
& 0.99793 && 0.680 && 1.395 && 0.320 && 0.420 && 1.240 &\\
& 0.99901 && 0.660 && 1.341 && 0.340 && 0.438 && 1.181 &\\
& 0.99960 && 0.640 && 1.292 && 0.360 && 0.455 && 1.129 &\\
& 0.99988 && 0.620 && 1.246 && 0.380 && 0.470 && 1.084 &\\
& 0.99998 && 0.600 && 1.205 && 0.400 && 0.484 && 1.045 &\\ \\
& 0.9978$^\star$ && 0.682 && 1.400 && 0.302 (0.318) && 0.419 && 1.246 &\\
& 0.9983$^\star$ && 0.674 && 1.379 && 0.308 (0.326) && 0.426 && 1.223 &\\
& 0.9983$^\dag$ && 0.674 && 1.379 && 0.309 (0.326) && 0.426 && 1.223 &\\
& 0.9986$^\dag$ && 0.669 && 1.365 && 0.315 (0.331) && 0.430 && 1.207 &\\
\hline
\end{tabular}
  \caption{Kerr space-time. $E_{\rm ISCO}$, $L_{\rm ISCO}$, maximum radiative efficiency in the NT model, angular frequency at the ISCO radius, and ISCO radius in Boyer-Lindquist coordinates for different value of the spin parameter. $^\star$ From Ref.~\cite{t}, without returning radiation. $^\dag$ From Ref.~\cite{li}, including returning radiation. See text for details. \label{tab}}
\end{table*}

\begin{figure*}
\includegraphics[type=pdf,ext=.pdf,read=.pdf,height=6cm]{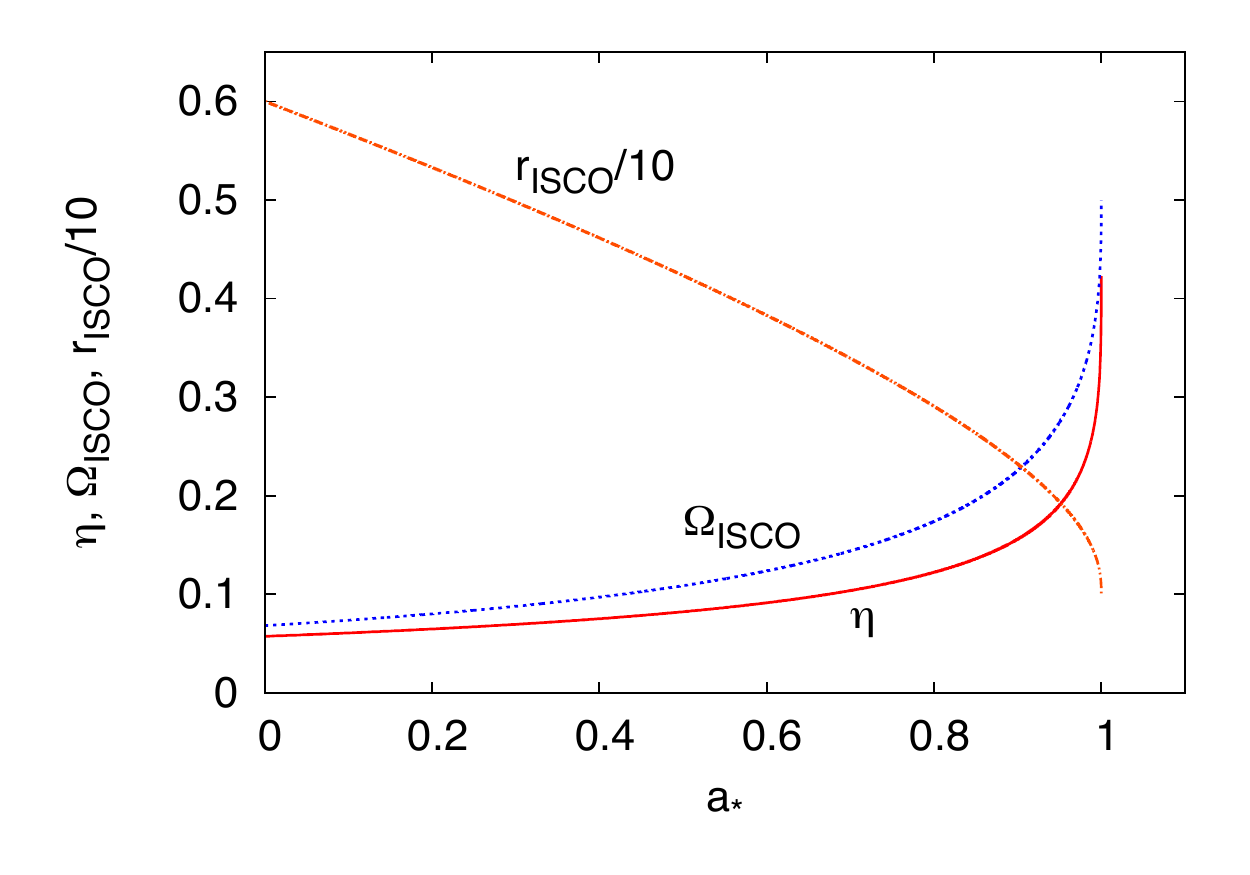}
\includegraphics[type=pdf,ext=.pdf,read=.pdf,height=6cm]{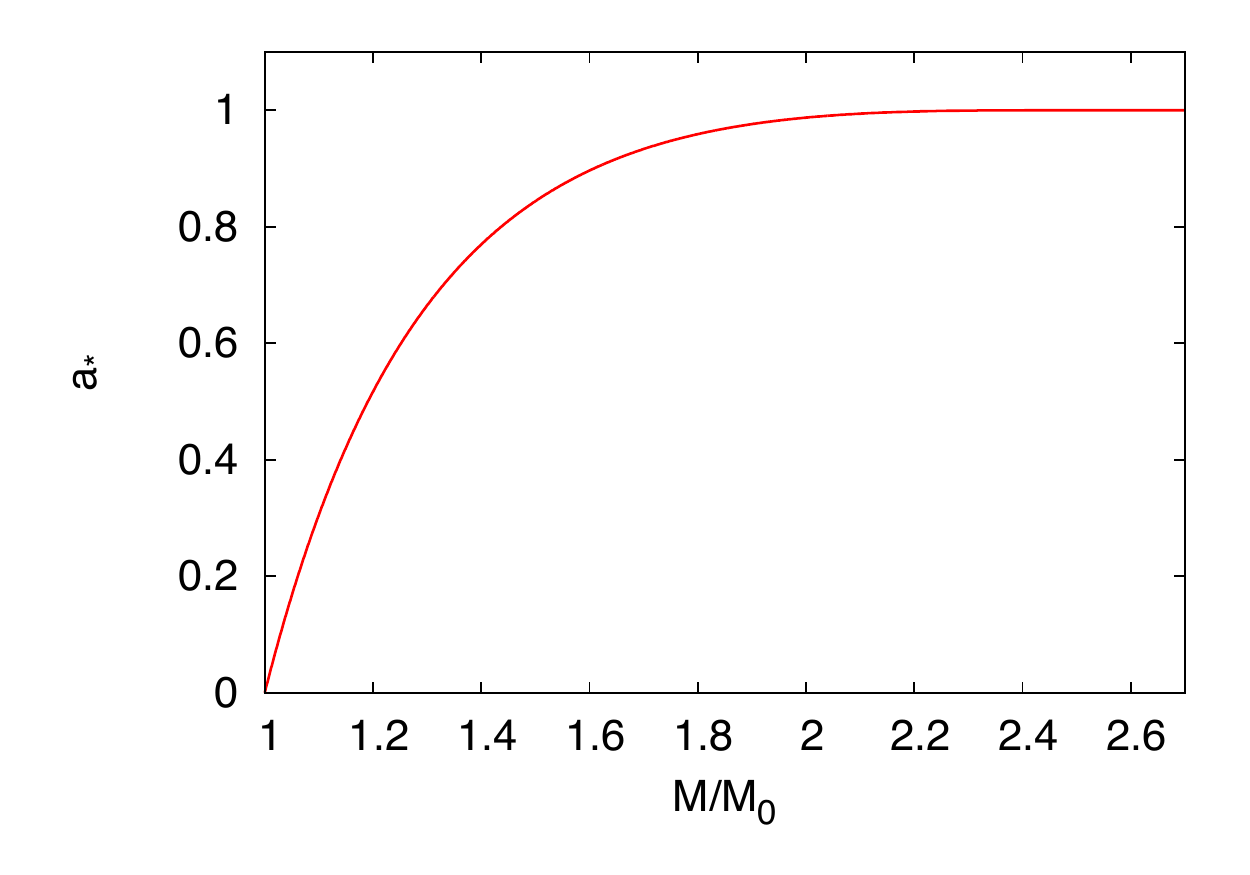}
\caption{Kerr space-time. Left panel: maximum radiative efficiency in the NT model, angular frequency at the ISCO radius, and ISCO radius (divided by 10) in Boyer-Lindquist coordinates as a function of the spin parameter. Units $M = 1$. Right panel: Evolution of the spin parameter for an initially non-rotating BH of mass $M_0$. See text for details.}
\label{f0}
\end{figure*}

\section{Kerr black holes} \label{s-k}

A Kerr BH is completely specified by two parameters: the mass, $M$, and the
spin parameter, $a_*$. In the NT model with $\eta_{\rm k} = 0$, the radiative 
efficiency of the accretion process $\eta$ is uniquely determined by the BH 
spin parameter, as $a_*$ sets the ISCO radius. There is a one-to-one 
correspondence between $a_*$ and $\eta$, and $\eta$ increases with 
increasing $a_*$\footnote{That is true only for Kerr BHs, i.e. when $a_* \le 1$.
It is not true in a Kerr background with arbitrary value of the spin 
parameter~\cite{th}.}. At least in principle, an estimate of $\eta$ could be used 
to infer the spin parameter of the BH. In practice, assuming the validity of
the NT model, we can deduce a lower bound for $a_*$, as in general
$\eta_{\rm k} \neq 0$.

The radiative efficiency in Kerr background for different values of the BH 
spin parameter is shown in Tab.~\ref{tab} and in the left panel of Fig.~\ref{f0}. 
If we use Eq.~(\ref{eq-eta}), for corotating accretion disks one finds that $\eta$ 
is in the range $\sim 0.057$ (Schwarzschild BH, $a_* = 0$) to $\sim 0.423$ 
(extreme Kerr BH, $a_* = 1$). However, $\eta$ increases slowly with $a_*$ 
for low values of the spin parameters and it increases much faster when 
$a_*$ approaches 1. For instance, $\eta > 0.30$ requires $a_* > 0.996$, 
$\eta > 0.34$ requires $a_* > 0.999$, and $\eta > 0.40$ demands $a_*$ 
extremely close to 1 (see Tab.~\ref{tab} and Fig.~\ref{f0}).

In the Kerr background, we can analytically integrate Eq.~(\ref{eq-a}) and
we obtain~\cite{b70}
\be
a_* &=& \sqrt{\frac{2}{3}}
\frac{M_0}{M} \left[ 4 - \sqrt{18 \frac{M_0^2}{M^2} - 2} \right] 
\quad ({\rm for} \; M/M_0 \le \sqrt{6}) \, ,
\nonumber\\
a_* &=& 1 \quad ({\rm for} \; M/M_0 > \sqrt{6}) \, ,
\ee
for an initially non-rotating BH with mass $M_0$. The equilibrium spin
parameter is thus $a_*^{\rm eq} = 1$, and it is reached after the object 
increased its mass by a factor $\sqrt{6} \approx 2.4$. The evolution of the
spin parameter as a function of $M/M_0$ is shown in Fig.~\ref{f0}, right
panel.

For an astrophysical BH, the situation is slightly different, as there is no
realistic mechanism to spin up the object to $a_*$ too close to 1.
The accretion process from a thin disk is still a very efficient mechanism,
but we have to include the effect of the radiation emitted by the disk and 
captured by the BH, Eqs.~(\ref{eq-eta2}) and (\ref{eq-a2}). In this case, 
we find slightly lower values for $\eta$ and $a_*^{\rm eq}$. However, as 
$\eta$ increases very quickly when $a_*$ approaches 1, even a tiny 
different in $a_*^{\rm eq}$ causes a non-negligible difference in the maximum 
value of $\eta$. As the radiation with angular momentum anti-parallel to the BH
spin has larger capture cross section, from Eqs.~(\ref{eq-a2}) one finds
$a_*^{\rm eq} < 1$. For instance, Thorne found $a_*^{\rm eq} = 0.9978$ when the
emission of the disk is isotropic, and $a_*^{\rm eq} = 0.9983$ when the
emission is limb-darkened~\cite{t}. The corresponding efficiencies [Eq.~(\ref{eq-eta2})]
are respectively 0.302 and 0.308. From Eq.~(\ref{eq-eta}), we would have 
found $0.318$ (for $a_* = 0.9978$) and $0.326$ (for $a_* = 0.9983$).
The effect of the returning radiation, which changes a little bit the emission
of the disk, introduces an even smaller correction~\cite{li}. The equilibrium 
spin parameter is now $a_*^{\rm eq} = 0.9983$ in the case of isotropic emission
and $a_*^{\rm eq} = 0.9986$ in the case of limb-darkened emission. The radiative
efficiencies turn out to be respectively 0.309 and 0.315 [0.326 and 0.331 if we
used Eq.~(\ref{eq-eta})]. In conclusion, if the super-massive BH candidates
are Kerr BHs, a realistic upper bound for $\eta$ should be around 0.32, as
these objects can unlikely be spun up to $a_*$ higher than roughly 0.998.

Tab.~\ref{tab} shows also the angular frequency at the ISCO, $\Omega_{\rm ISCO}$,
and the value of the ISCO radius in Boyer-Lindquist coordinates, $r_{\rm ISCO}$,
for the same values of the spin parameter. Even for $\Omega_{\rm ISCO}$ and 
$r_{\rm ISCO}$ there is a one-to-one correspondence (for a given mass $M$)
with the spin parameter of the BH. In the case of corotating disks, the angular 
frequency at the ISCO increases with $a_*$, from $\sim$0.068~$M$ ($a_* = 0$) 
to 0.5~$M$ ($a_* = 1$). The ISCO radius decreases with $a_*$, from 6~$M$ 
($a_* = 0$) to $M$ ($a_* = 1$). $\Omega_{\rm ISCO}$ and $r_{\rm ISCO}$
as a function of the spin parameter are shown (for corotating disk) in the left 
panel of Fig.~\ref{f0}.

\begin{figure*}
\includegraphics[type=pdf,ext=.pdf,read=.pdf,height=6cm]{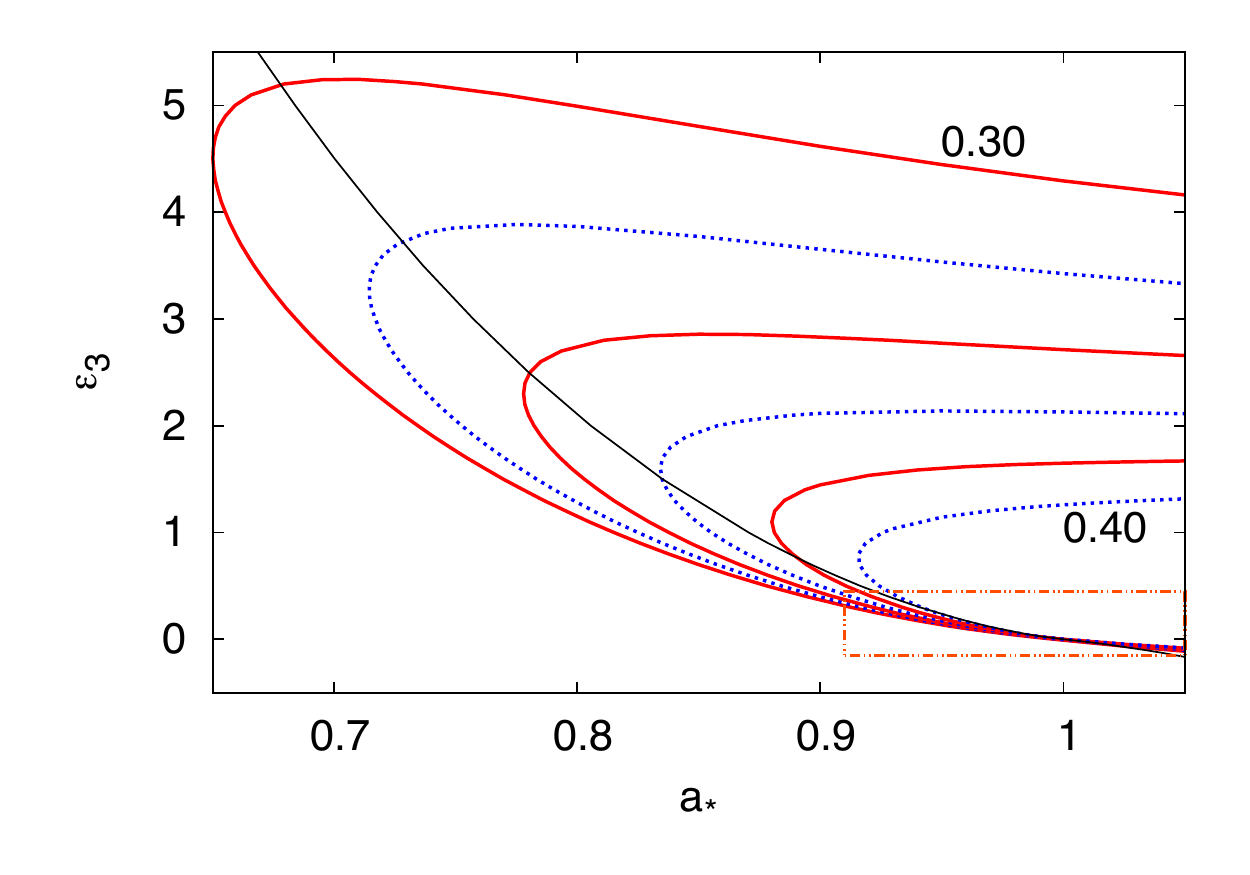}
\includegraphics[type=pdf,ext=.pdf,read=.pdf,height=6cm]{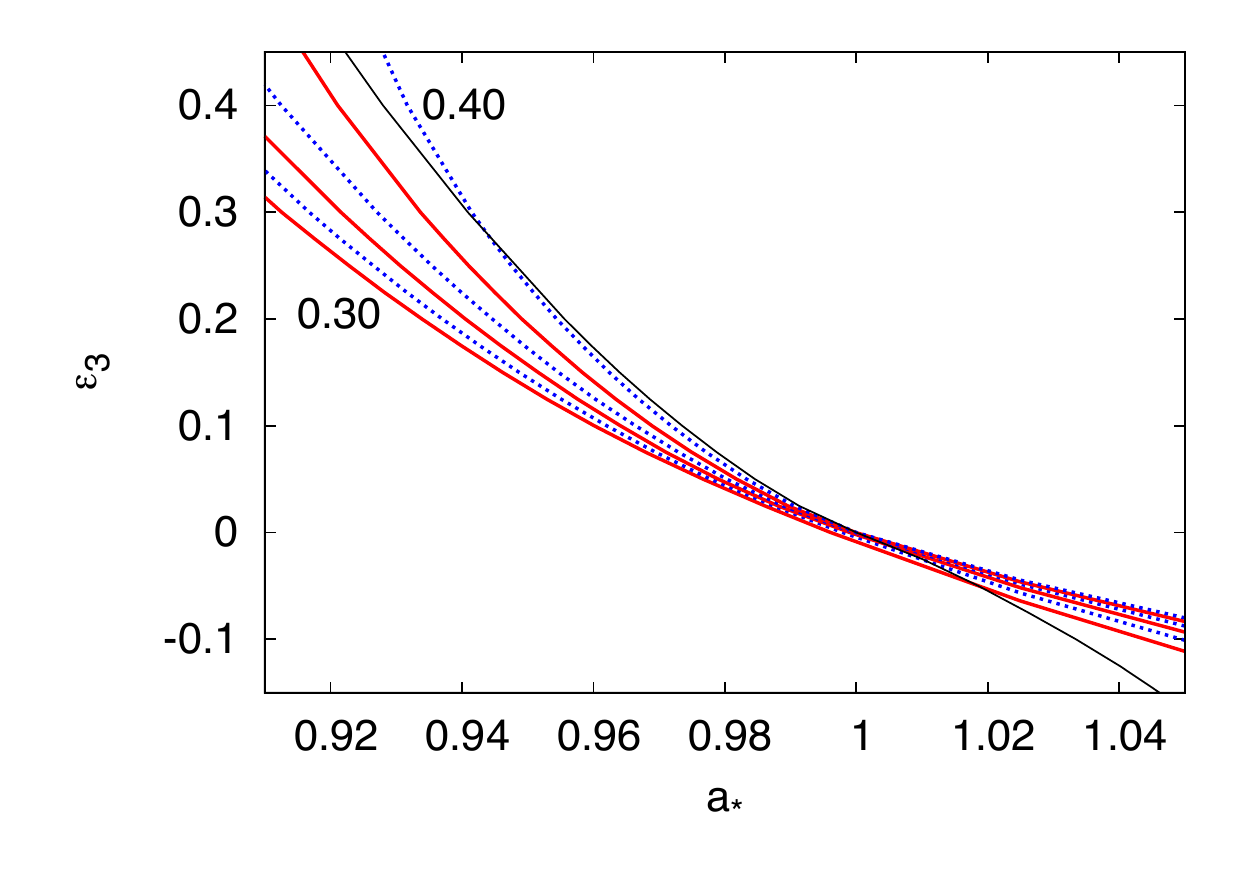}
\caption{JP space-time with deformation parameter $\epsilon_3$
and $\epsilon_i = 0$ for $i \neq 3$. Contour plots of the radiative
efficiency $\eta = 1 - E_{\rm ISCO}$: $\eta = 0.30$ (red solid curve),
0.32 (blue dotted curve), 0.34 (red solid curve), 0.36 (blue dotted 
curve), 0.38 (red solid curve), 0.40 (blue dotted curve). The black
solid curve is the equilibrium spin parameter $a_*^{\rm eq}$ as 
inferred from Eq.~(\ref{eq-a}) and represents the maximum value
for the spin parameter of the super-massive BH candidates. This
means that the region with $a_* > a_*^{\rm eq}$ is not allowed. 
The right panel is simply the enlargement of the area inside the 
orange box in the left panel.}
\label{f1}
\end{figure*}

\begin{figure*}
\includegraphics[type=pdf,ext=.pdf,read=.pdf,height=6cm]{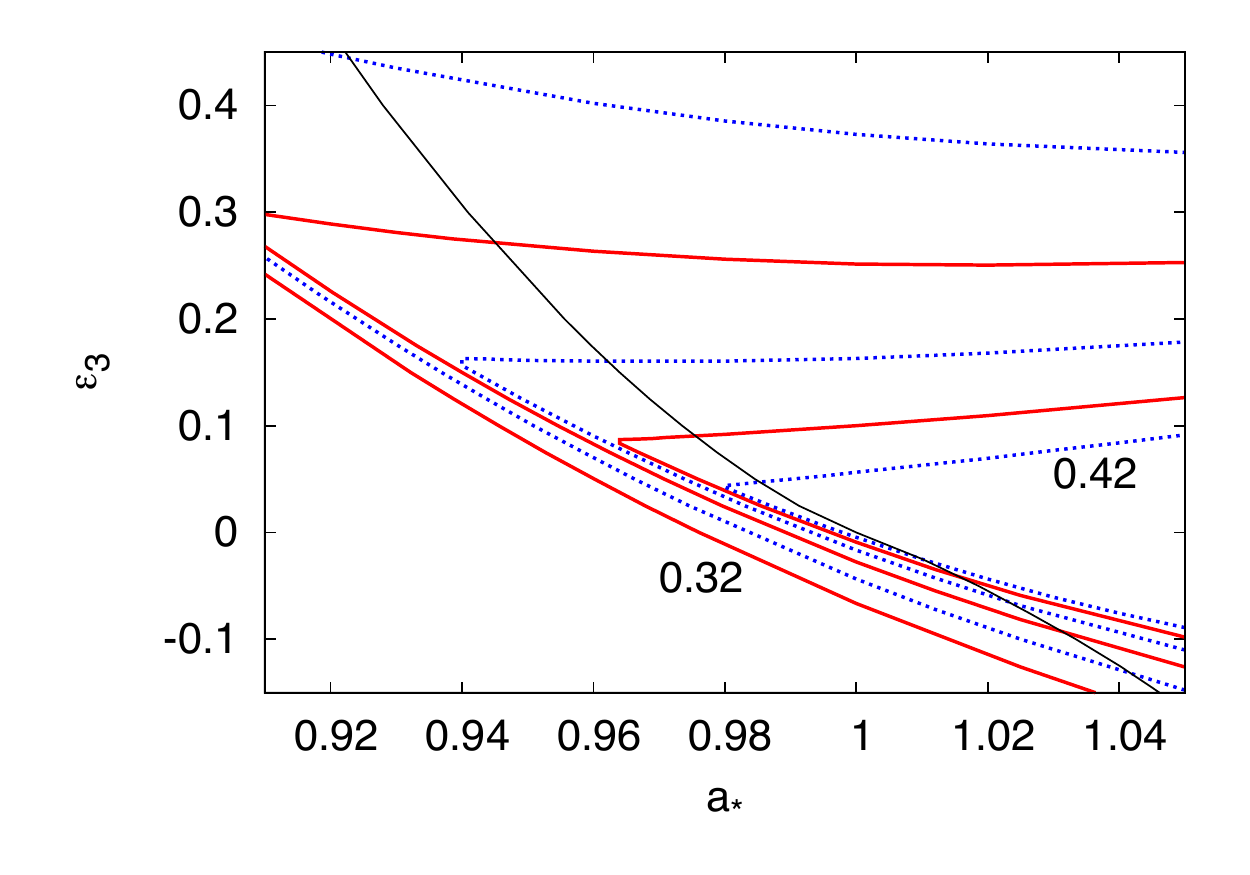}
\includegraphics[type=pdf,ext=.pdf,read=.pdf,height=6cm]{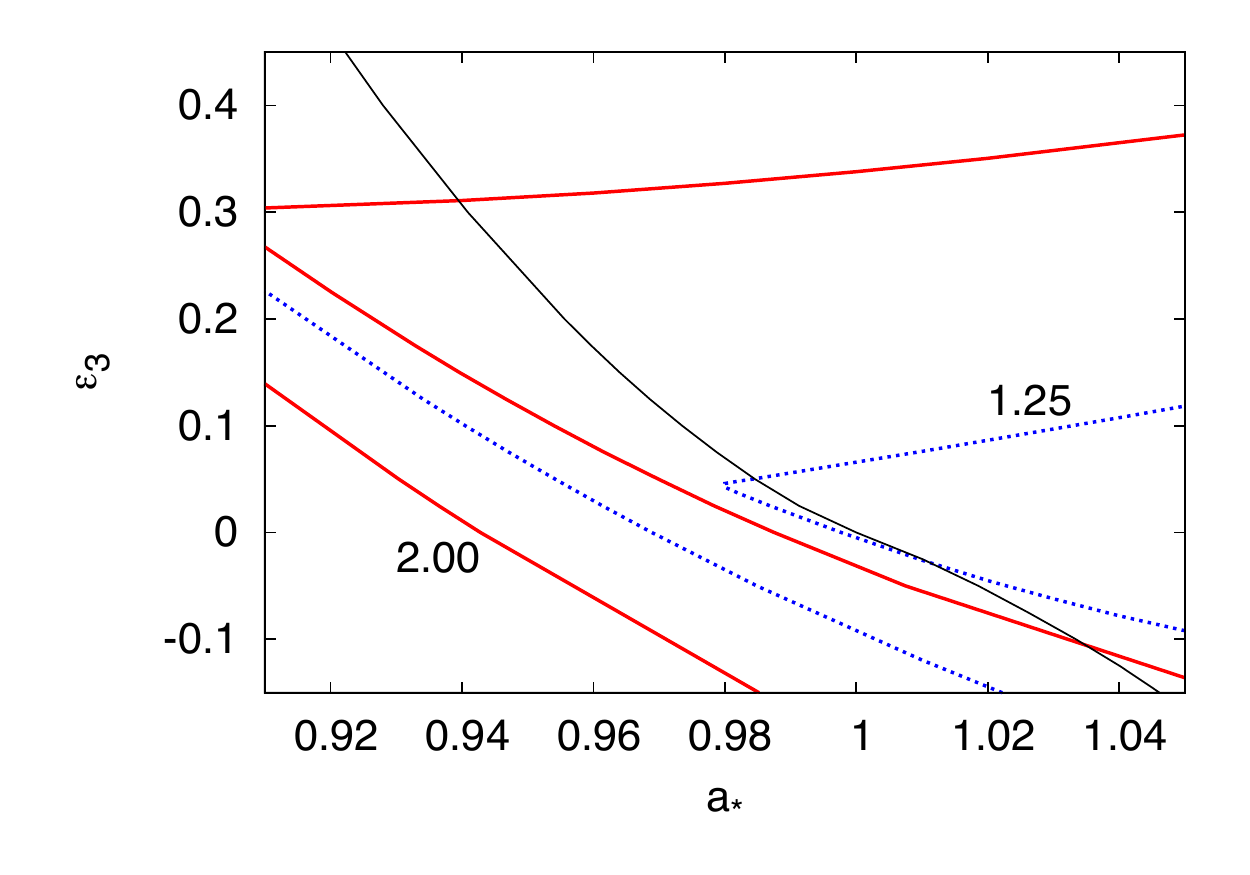}
\caption{JP space-time with deformation parameter $\epsilon_3$
and $\epsilon_i = 0$ for $i \neq 3$. Left panel: contour plot of 
the angular frequency at the ISCO radius: $\Omega_{\rm ISCO}/
M = 0.32$ (red solid curve), 0.34 (blue dotted curve), 0.36 (red 
solid curve), 0.38 (blue dotted curve), 0.40 (red solid curve), 0.42 
(blue dotted curve). Right panel: contour plot of the ISCO radius
in Boyer-Lindquist coordinates: $r_{\rm ISCO}/M = 2.00$ (red solid 
curve), 1.75 (blue dotted curve), 1.50 (red solid curve), 1.25 (blue 
dotted curve). The black solid curve is the equilibrium spin parameter 
$a_*^{\rm eq}$ as inferred from Eq.~(\ref{eq-a}).}
\label{f2}
\end{figure*}

\begin{figure*}
\includegraphics[type=pdf,ext=.pdf,read=.pdf,height=6cm]{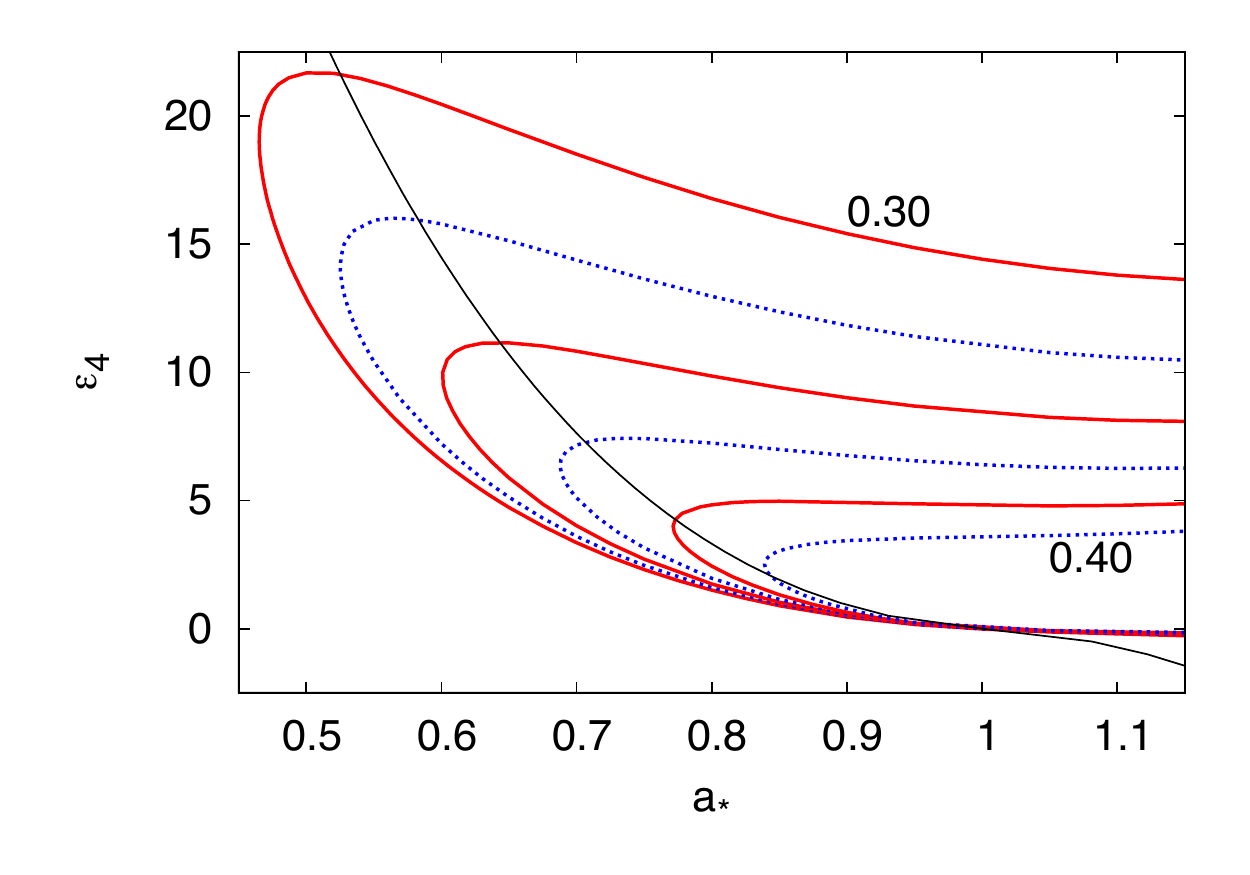}
\includegraphics[type=pdf,ext=.pdf,read=.pdf,height=6cm]{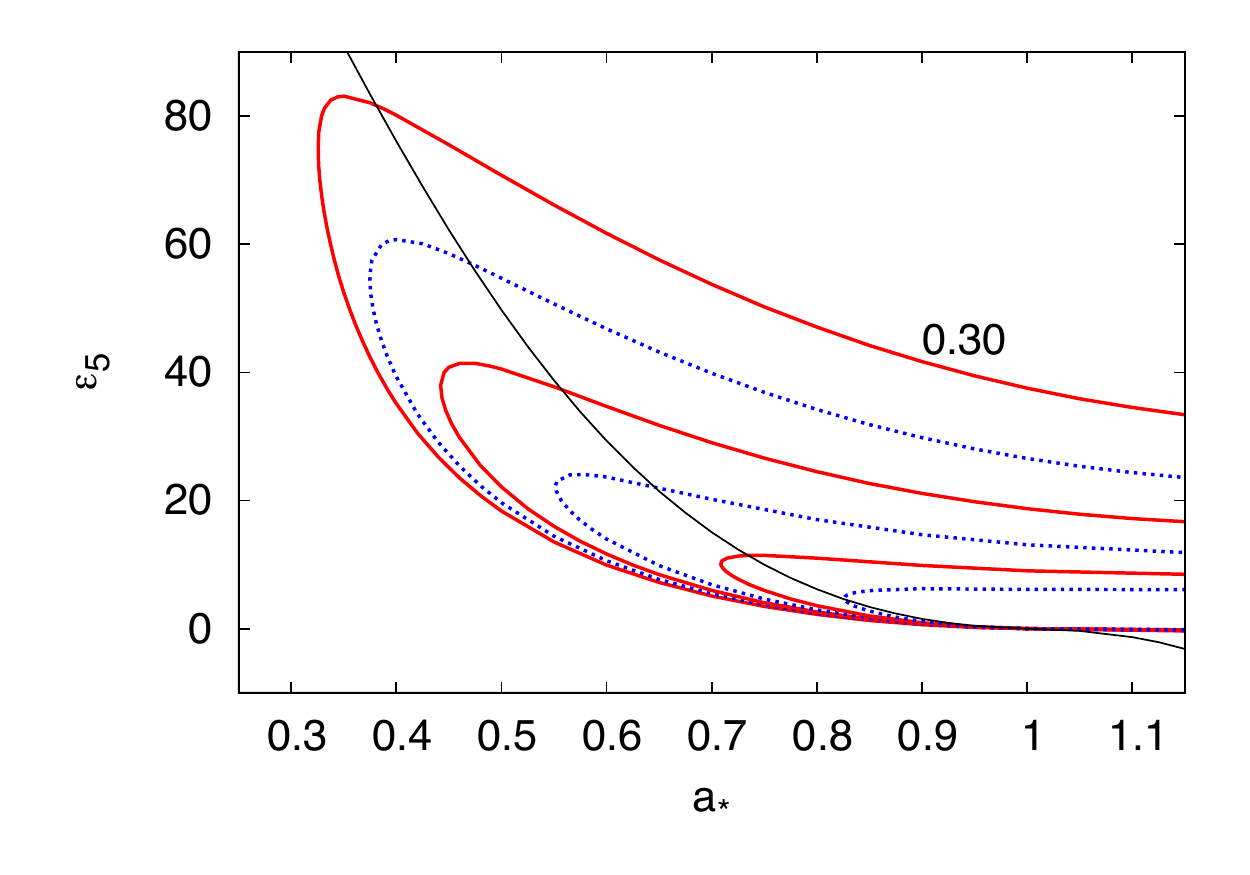}
\caption{As in Fig.~\ref{f1}, for the JP space-time with deformation 
parameter $\epsilon_4$ and $\epsilon_i = 0$ for $i \neq 4$ 
(left panel) and for the one with deformation parameter $\epsilon_5$ and 
$\epsilon_i = 0$ for $i \neq 5$ (right panel).}
\label{f3}
\end{figure*}

\section{Johannsen-Psaltis space-times}\label{s-jp}

The JP space-times have been proposed in~\cite{jp} explicitly to be used
to test the Kerr geometry around astrophysical BH candidates. They
describe BHs (at least for $a_*$ lower than a critical value) and they are 
an extension of the Kerr solution, in the sense that here the compact
objects are specified by the mass, the spin angular momentum, and an
infinite number of deformation parameters $\epsilon_i$ ($i = 3$, 4, 5, etc.) 
measuring deviations from the Kerr geometry. When all the deformation 
parameters vanish, one recovers exactly the Kerr background. The explicit 
expression of the metric is reported in Appendix~\ref{a-1}. The JP metric 
is not a solution of any known gravity theory, but it seems to be a very 
convenient framework to test the Kerr BH hypothesis.

The NT model for thin accretion disks can be easily extended to non-Kerr 
backgrounds. The first important difference is the determination of
the ISCO radius. In the Kerr background, circular orbits on the equatorial
plane are always vertically stable and the ISCO is set by the orbital
stability along the radial direction. If the compact object is more oblate 
than a Kerr BH, that remains true. On the contrary, if the compact object
is more prolate (including the case in which the object is still oblate, but simply 
less oblate than a Kerr BH with the same spin) both kinds of instabilities are possible.
When the ISCO radius is marginally stable along the radial direction,
as in Kerr, it is at the minimum of the energy of equatorial circular orbits.
When instead the ISCO is marginally stable along the vertical direction,
such a minimum does not exist in general. In the latter case, the curve
of the energy of equatorial circular orbits around the ISCO is clearly steeper,
causing a lower value of $E_{\rm ISCO}$ with respect to the case of
radially unstable ISCO with the same radius. The total efficiency $\eta$ 
is therefore higher.

As Eqs.~(\ref{eq-a}) and (\ref{eq-a2}) depend on the geometry of the
space-time, the equilibrium spin parameter in these metrics is
determined by the deformation parameters $\epsilon_i$. We can 
notice that in the case of the super-massive BH candidates, 
$a_*^{\rm eq}$ should be the maximum value for
the spin parameter of these objects, independently of their actual
nature~\cite{b5,b1,b3}. Indeed, for the super-massive BH candidates
the initial value of $a_*$, i.e. the one at the time
of their birth, is thought to be completely irrelevant, as the mass
of these objects has increased by several orders of magnitude from 
the original one and the spin has evolved accordingly. Long term
accretion from a thin disk can efficiently spin the BH up to $a_*^{\rm eq}$
(but it spins the BH down if $a_* > a_*^{\rm eq}$), while other processes 
(chaotic accretions, minor and major mergers) more likely spin the object 
down to $a_* \sim 0$. So, the value of $a_*^{\rm eq}$ we can infer
from Eqs.~(\ref{eq-a}) and (\ref{eq-a2}) can be used to exclude the
region $a_* > a_*^{\rm eq}$ in the diagrams spin parameter vs deformation
parameter. This point will be crucial for what follows.

Let us now start considering the properties of the JP space-time
with deformation parameter $\epsilon_3$ and $\epsilon_i = 0$ for 
$i \neq 3$. Fig.~\ref{f1} shows the efficiency $\eta$, as deduced
from Eq.~(\ref{eq-eta}). In particular, we are interested in the most
efficient systems and I report the curves on the $a_*\epsilon_3$-plane
with $\eta = 0.30$, 0.32, 0.34, 0.36, 0.38, 0.40. The black solid 
curve marks the equilibrium spin parameter $a_*^{\rm eq}$. For 
$\epsilon_3 > 0$, the compact object is more prolate than Kerr
and $a_*^{\rm eq} < 1$. For $\epsilon_3 < 0$, the object is more oblate
and $a_*^{\rm eq} > 1$. It is thus clear that observations can potentially
provide very strong constraints for $\epsilon_3 < 0$, and much weaker
bounds for $\epsilon_3 > 0$. In particular, for $\epsilon_3 > 0$,
the maximum value of $a_*$ may be even significantly lower than 1,
but still $\eta$ can be very high. Interestingly, for $\epsilon_3 \neq 0$,
$\eta$ does not increases abruptly as $a_*$ approaches $a_*^{\rm eq}$
and therefore the effect of the radiation emitted by the disk and captured 
by the BH may be neglected. Indeed, we can assume that the radiation
captured by the BH decreases $a_*^{\rm eq}$ by $\sim 0.002$, as in
Kerr (actually the correction to $a_*^{\rm eq}$ should be smaller than 
0.002, as the ISCO radius is larger, see the right panel in Fig.~\ref{f2}). 
This is equivalent to replace the curve of $a_*^{\rm eq}$ in Fig.~\ref{f1}
with the curve $a_*^{\rm eq} - 0.002$. At this point, we have to decrease
$\eta$ by $\sim 0.01$. Except for the Kerr space-time and for the space-times
with small deviations from the Kerr geometry, the two effects should
not cause significant changes. In particular, we still find a large region
with objects with $\eta > 0.32$, which would be impossible for a Kerr
BH in a realistic astrophysical context.

Fig.~\ref{f2} shows the contour plot of the angular frequency at the
ISCO, $\Omega_{\rm ISCO}$, (left panel) and the one of the ISCO
radius in Boyer-Lindquist coordinates, $r_{\rm ISCO}$, (right panel). While accretion
in JP space-time may have high efficiency, thus mimicking a Kerr BH 
with $a_*$ extremely close to 1 (or even exceeding the efficiency of
a Kerr BH with $a_* = 1$), the same object would have $\Omega_{\rm ISCO}$ 
and $r_{\rm ISCO}$ more similar to the one expected for Kerr BHs 
with lower spin parameter. $\Omega_{\rm ISCO}$ can be potentially 
inferred in several ways from observations (e.g. from the variability of 
the source) and that may be useful to break the degeneracy between 
$a_*$ and $\epsilon_3$ in $\eta$.

Lastly, we can consider JP backgrounds with a different deformation 
parameter. Interestingly, one finds essentially the same picture. In
Fig.~\ref{f3}, I report the contour plots of $\eta$ for the JP space-time
with deformation parameter $\epsilon_4$ and $\epsilon_i = 0$ for
$i \neq 4$ (left panel) and for the one with deformation parameter
$\epsilon_5$ and $\epsilon_i = 0$ for $i \neq 5$ (right panel). For
these two cases, even the contour plots of $\Omega_{\rm ISCO}$ and 
$r_{\rm ISCO}$ (not shown here) are fairly similar to the ones in
Fig.~\ref{f2}.

\begin{figure*}
\includegraphics[type=pdf,ext=.pdf,read=.pdf,height=6cm]{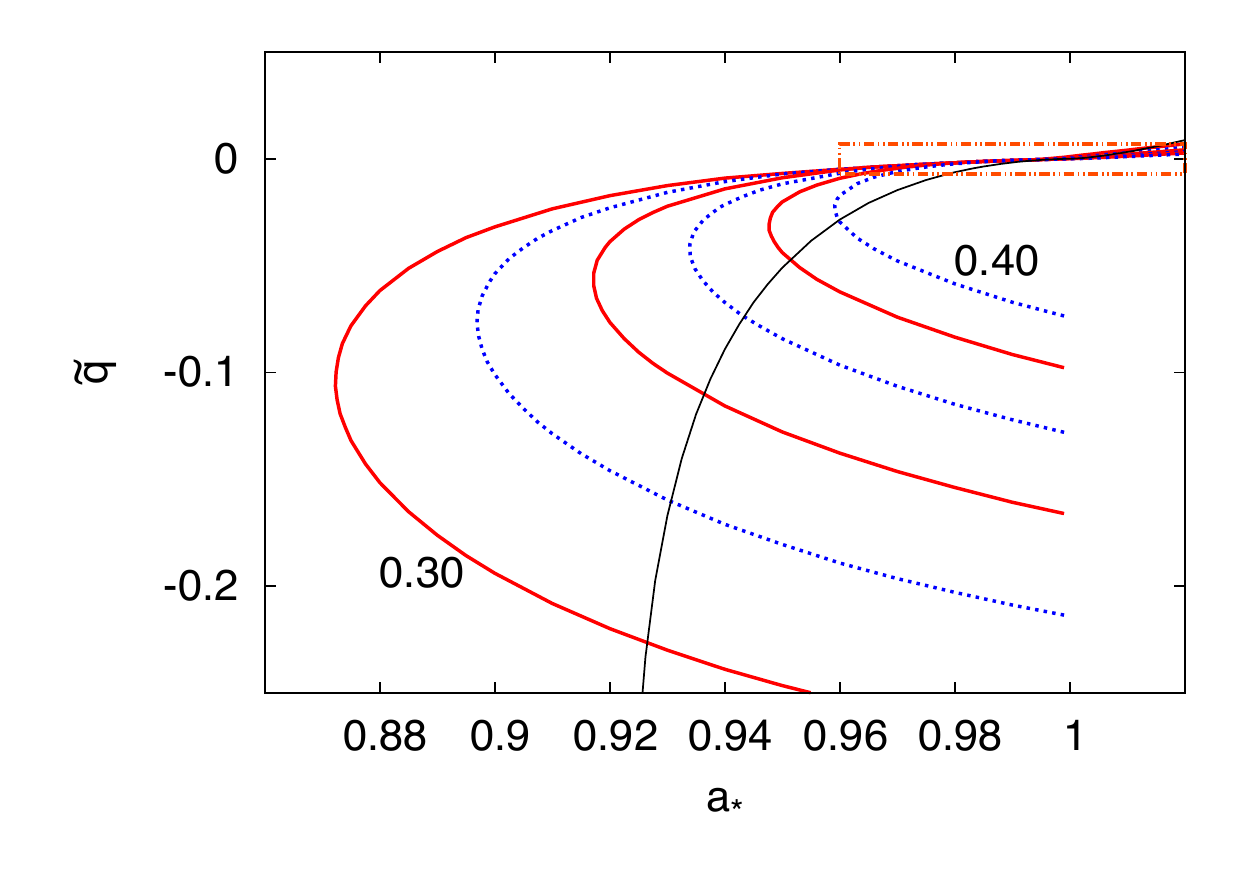}
\includegraphics[type=pdf,ext=.pdf,read=.pdf,height=6cm]{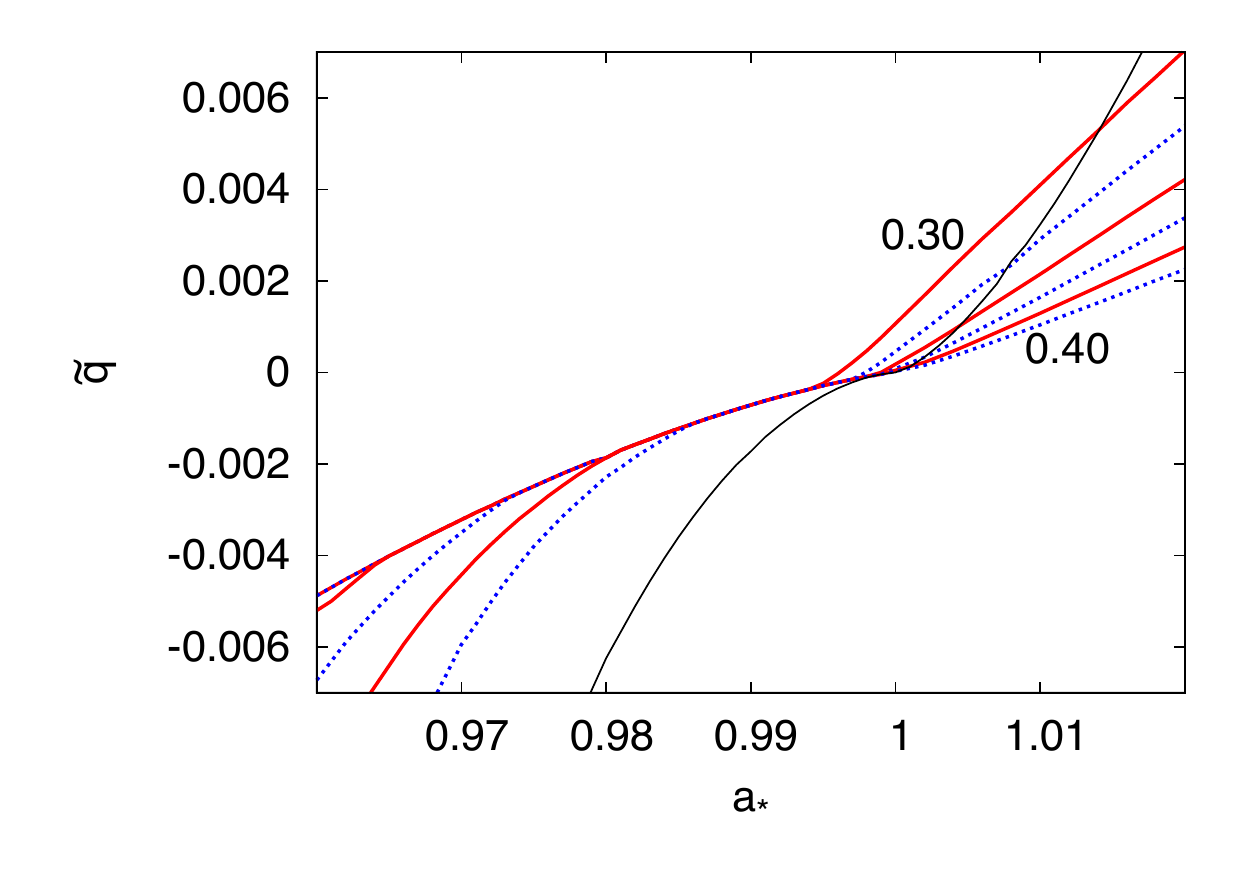}
\caption{As in Fig.~\ref{f1} for the MN space-time with deformation 
parameter $\tilde{q}$. The compact object is oblate (prolate) when 
$\tilde{q} > - 1$ ($\tilde{q} < - 1$), but it is more oblate (prolate) 
than a Kerr BH with the same value of the spin parameter for 
$\tilde{q} > 0$ ($\tilde{q} < 0$).}
\label{f4}
\end{figure*}

\begin{figure*}
\includegraphics[type=pdf,ext=.pdf,read=.pdf,height=6cm]{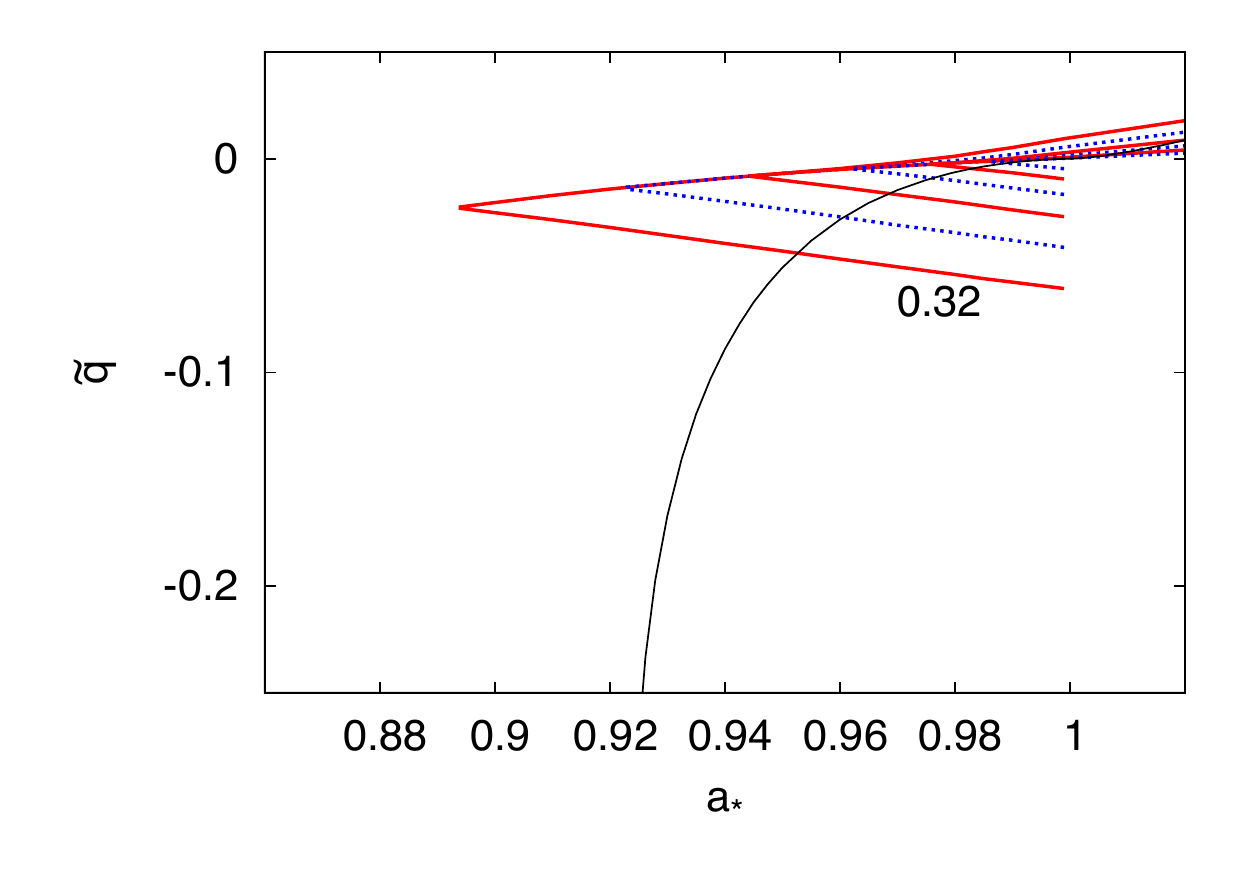}
\includegraphics[type=pdf,ext=.pdf,read=.pdf,height=6cm]{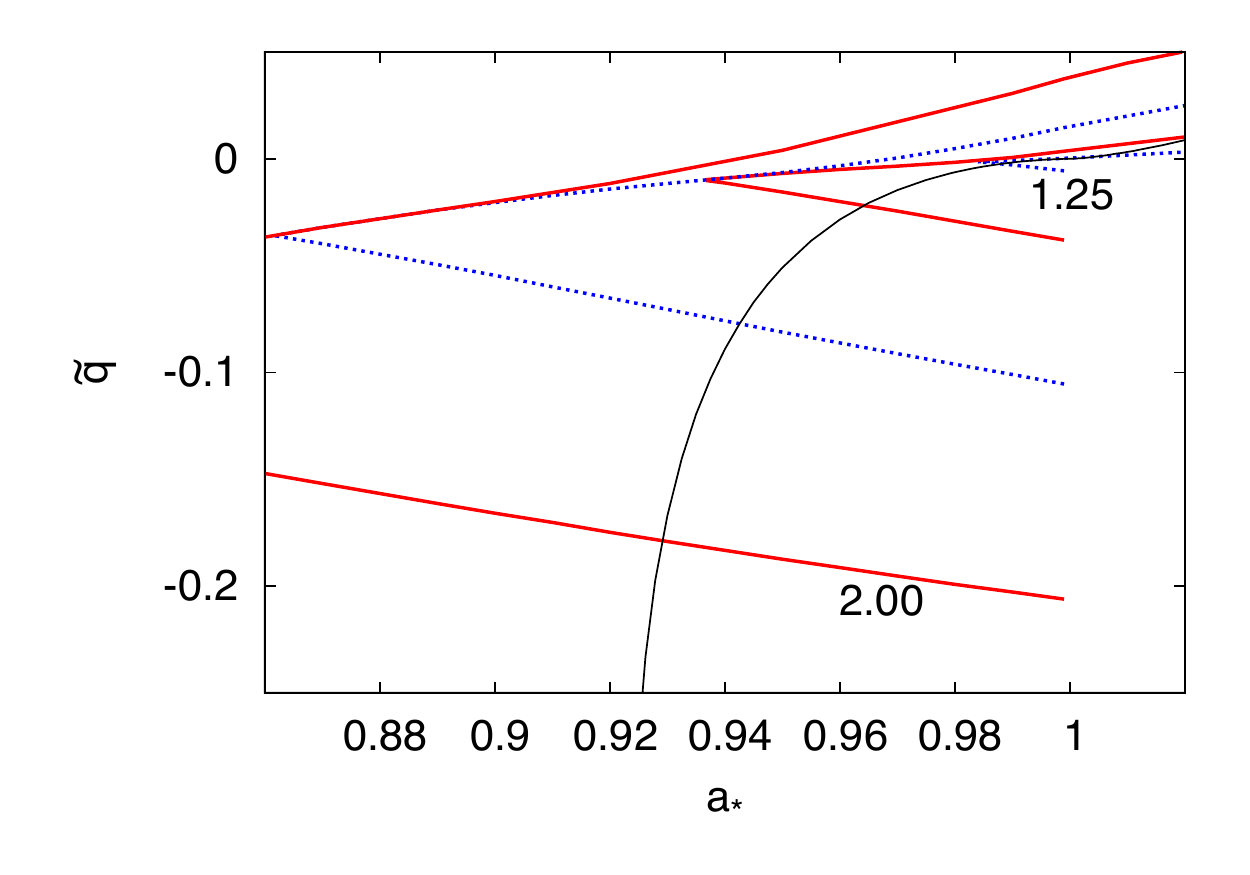}
\caption{As in Fig.~\ref{f2} for the MN space-time with deformation 
parameter $\tilde{q}$.}
\label{f5}
\end{figure*}

\section{Manko-Novikov space-times}\label{s-mn}

The MN space-times are stationary, axisymmetric, and asymptotically
flat exact solutions of the Einstein's vacuum equations~\cite{mn}.
These metrics can describe the exterior gravitational field of a compact
object with mass $M$, spin parameter $a_*$, and arbitrary mass-multipole
moments, while the current-multipole moments are fixed by the former.
The simplest non-Kerr object has three free parameters: mass $M$, spin
parameter $a_*$, and anomalous quadrupole moment $\tilde{q}$
(see Appendix~\ref{a-2}). $\tilde{q}$ is defined by
\be\label{e-qq}
Q = (1 + \tilde{q}) Q_{\rm Kerr} \, ,
\ee
where $Q$ is the mass-quadrupole moment of the compact object,
while $Q_{\rm Kerr} = - a_*^2 M^3$ is the one of a Kerr BH with the
same spin parameter $a_*$. The case $\tilde{q}=0$ corresponds to the Kerr
metric, while for $\tilde{q} > 0$ ($\tilde{q} < 0$) the object is more oblate
(prolate) than a Kerr BH. The properties of this subclass of MN space-times
with deformation parameter $\tilde{q}$ were studied in~\cite{glm} and~\cite{bb2}. 
When $\tilde{q} < 0$, for a quite restricted set of values of $a_*$ and $\tilde{q}$, the 
plunging region does not connect the ISCO to the compact object~\cite{bb2}. That 
occurs when inside the ISCO there is another region with stable circular orbits 
on the equatorial plane and energy lower than $E_{\rm ISCO}$. In these
space-times, the gas's particles plunging from the ISCO get trapped between 
the ISCO and the compact object. As the gas need to radiate additional energy 
and angular momentum in order to fall to the compact object, the value of
the radiative efficiency turns out to be higher than the one inferred from 
Eq.~(\ref{eq-eta}) or Eq.~(\ref{eq-eta2}). The actual value of $\eta$ depends
on the astrophysical processes. For instance, Ref.~\cite{bb2} discussed a simple 
model in which the gas forms a thick disk inside the ISCO and $\eta = 1 - 
E_{\rm in}$, where $E_{\rm in}$ is the specific energy at the inner edge of the 
thick disk, is up to a few percent higher than the one predicted by
Eq.~(\ref{eq-eta}).

Fig. ~\ref{f4} shows the radiative efficiency $\eta$ as computed from Eq.~(\ref{eq-eta}).
The MN metric is written in prolate spheroidal coordinates, which require 
$a_* < 1$; here the region $a_* \ge 1$ is covered by the Malko-Mielke-Sanabria
G{\'o}mez solution~\cite{mms1} which can be parametrized by the same three
parameters of our subclass of the MN metric (i.e. $M$, $a_*$, and $\tilde{q}$) and was
shown to be very similar to the MN space-time in the common region of 
validity~\cite{b3}. It is clear that there are strong analogies with the JP space-times. 
For objects more oblate than Kerr, $a_*^{\rm eq} > 1$ and a high radiative efficiency 
is possible only for small deviations from the Kerr geometry. For objects more 
prolate than Kerr, $a_*^{\rm eq} < 1$ and when the ISCO radius is set by the
orbital stability along the vertical direction $\eta$ can be high even for objects
rotating slower. Fig.~\ref{f5} shows the angular frequency at the ISCO, 
$\Omega_{\rm ISCO}$, and the ISCO radius in Boyer-Lindquist coordinates, 
$r_{\rm ISCO}$. As in the JP background, the same object with high radiative efficiency 
has lower $\Omega_{\rm ISCO}$ and larger $r_{\rm ISCO}$. The case of the
MN space-time with $\tilde{q} = 0$ and an anomalous mass-hexadecapole 
moment $\tilde{h}$ shows similar properties in $\eta$, $\Omega_{\rm ISCO}$, and 
$r_{\rm ISCO}$.

\section{Discussion}\label{s-d}

The radiative efficiency of a source can be deduced from the formula
$\eta = L_{\rm bol}/\dot{M}$. In general, however, it is not easy to get an
estimate of $\eta$, because a measurement of the mass accretion rate 
can be problematic. The Soltan's argument provides an elegant way
to determine the mean radiative efficiency of AGN, $\bar{\eta}$, from
the mean BH mass density in the contemporary Universe and the AGN
luminosity per unit volume integrated over time~\cite{s}. There are several
sources of uncertainty in the final result, but a conservative bound seems
to be $\bar{\eta} > 0.15$~\cite{erz}. If the compact objects in AGN are 
Kerr BHs, this constraint requires $\bar{a}_* > 0.89$. In Ref.~\cite{ho}, the
authors used a revised version of the Soltan's argument, in which some
important assumptions are not necessary. They found a mean radiative 
efficiency $\bar{\eta} \approx 0.30 - 0.35$ in the redshift interval $0.4 < z < 2.1$.

Recently, it has been proposed a way to estimate the radiative efficiency of
individual AGN~\cite{dl}. The mass accretion rate $\dot{M}$ can indeed be 
determined from the low frequency region of the thermal spectrum of the 
accretion disk of these objects, if the mass $M$ is known. The standard 
accretion disk model cannot reproduce the observed spectrum of AGN. That 
is due to reprocessing, Comptonization in a corona, radiative transfer effects 
in the inner accretion disk, an so on. Here the key-point is that this method 
relies on the validity of the simple thin-disk model at relatively large radii, 
where these effects are thought to be irrelevant. One can then estimate the 
total luminosity of the source, irrespective of its exact production mechanism, 
and infer the radiative efficiency.
So, the redistribution of the radiation emitted in the accretion process does 
not affect the measurement of radiative efficiency; for more details, see
Sections 2.3 and 4.3 of Ref.~\cite{dl}. The authors found a strong correlation 
of $\eta$ with $M$, raising from $\eta \sim 0.03$ when $M \sim 10^7$~$M_\odot$ and 
$L_{\rm bol}/L_{\rm Edd} \sim 1$ to $\eta \sim 0.4$ when $M \sim 10^9$~$M_\odot$ 
and $L_{\rm bol}/L_{\rm Edd} \sim 0.3$.

The most massive objects in AGN with $M \sim 10^9$~$M_\odot$ may
be excellent candidates to test the Kerr BH hypothesis. The standard 
criterion to consider thin disks is indeed that $L_{\rm bol}/L_{\rm Edd}$
does not exceed 0.3, corresponding to an opening angle $h$ not larger
than 0.05. In this case, the estimate of $\eta$ may provide
a measurement of the specific energy at the ISCO (or at least a lower bound,
as we are neglecting $\eta_{\rm k}$ and $\zeta_{\rm E}$). The estimate of
$\eta$ can then be used to constrain possible deviations from the Kerr
geometry and may have some advantages with respect to the most popular
techniques to measure the BH spin parameter, i.e. the continuum fitting
method and the K$\alpha$ iron line analysis. The continuum fitting method 
can be used only for stellar-mass BH candidates, for which there is no 
argument to constrain $a_*$: the value of the spin parameter of the BH 
candidates in X-ray binary systems should
reflect the one at the time of the formation of the object and it is definitively
impossible to predict its value relaxing the assumption that GR
is the correct theory of gravity. The method of the K$\alpha$ iron line can 
be used for the super-massive BH candidates in galactic nuclei (so we can
impose $a_* < a_*^{\rm eq}$), but it necessarily relies on an 
ad hoc intrinsic surface brightness profile and there is no way to determine 
it a priori. Although simulations have the potential to compute the surface 
brightness profile, the standard required to test the Kerr background is a 
much higher level of reliability than is likely to be met at any time in the 
foreseeable future.

Let us now assume that future GRMHD simulations confirm the validity
of the NT model for thin accretion disks and that future observations provide 
robust evidence that the most massive objects in AGN have a very high radiative
efficiency and a moderate mass accretion rate. If we exclude the possibility 
of objects more prolate than Kerr, which may be difficult to explain 
theoretically\footnote{The true problem is not the existence of compact objects
less oblate than a Kerr BH with the same spin parameter, but that the objects
we are talking about have the ISCO marginally stable along the vertical
direction. In the case of the JP metric, these objects are BHs with two disconnected
event horizons, one above and one below the equatorial plane~\cite{bm}.
These objects more prolate than Kerr and high radiative efficiency have
therefore ``two centers of attraction'', one above and one below the equatorial 
plane. It is not clear at all if such a system can exist and be stable.}, 
from the right panel of Fig.~\ref{f4} we can see that $\tilde{q}$ can be constrained
at the level of $\sim 10^{-3}$. That roughly corresponds to test the mass-quadrupole 
moment of these objects with a precision $\sim 10^{-3}$, see Eq.~(\ref{e-qq}). 
For a self-gravitating fluid with reasonable equations of state, $\tilde{q}$ is 
much larger, i.e. the compact object becomes significantly more oblate than a
Kerr BH when the spin parameter increases. For instance, a neutron star
should have $\tilde{q} > 1$~\cite{lp}. A similar bound may 
be obtained by observing the gravitational waves emitted by an EMRI
with future space-based gravitational-wave detectors
(see the third reference in~\cite{r}), even if a direct comparison is not
possible, as here we are probing the geometry of the space-time very
close to the compact objects, where higher orders multipole moments
are not really negligible, while gravitational wave detectors will study the 
space-time at larger distances, and therefore they will be sensitive to
the mass-quadrupole moment only. 
If we do not exclude a priori the existence of compact objects more
prolate than a Kerr BH, the constraint is much weaker: for $\eta > 0.30$,
deviations from the Kerr mass-quadrupole moments can still be up to
20\%. Let us now consider a more exotic situation: the NT model works and 
observations find unambiguously $\eta > 0.32$. In absence of torque at the 
inner edge of the disk, it would be difficult to explain this outcome in the 
Kerr background, and new physics may be invoked. Let us notice, however, 
that a non-Kerr background can explain high $\eta$, say up to $\eta \sim 0.4$, 
but higher values may not be natural, as we still have $a_* < a_*^{\rm eq}$.

\section{Conclusions}\label{s-c}

There is some evidence that the most massive BH candidates in AGN have a 
high radiative efficiency and a thin accretion disk. In this case, they could be excellent
candidates to test GR in the strong field regime and, in particular, the Kerr BH
paradigm. A robust observation of a high radiative efficiency together with the
confirmation of the validity of the standard accretion disk model for the accretion
process onto these objects would constrain possible deviations from the Kerr
geometry of the space-time around astrophysical BH candidates. If we restrict
our attention only to objects more oblate than a Kerr BH with the same spin
parameter, a radiative efficiency $\eta > 0.30$ requires that the quadrupole
moment of the compact object deviates not more than $\sim 10^{-3}$ with respect
to the one of a Kerr BH. If we allow for the existence of compact objects more
prolate than a Kerr BH (whose existence, however, may be questionable,
see the footnote~3), the constraints is much weaker, at the level of 20\%.
On the other hand, the observation of an accreting BH candidate with $\eta > 0.32$
might indicate that the super-massive BH candidates in galactic nuclei are
not the Kerr BH of GR, as there is no realistic astrophysical mechanism capable
to spin up them to $a_* > 0.998$.

For the time being, there are at least two main issues to address before using
the measurement of the radiative efficiency of individual AGN to test GR:
\begin{enumerate}
\item We do not really know if the NT model can describe thin accretion disks
around astrophysical BH candidates. If it can or it is easy to estimate the necessary
corrections, the measurement of $\eta$ could provide an interesting way to
probe the geometry of the space-time around BH candidates. If it cannot,
it seems to be impossible to test GR (with this or other approaches) from the
properties of the radiation emitted by the gas in the accretion disk and we should
wait for the advent of gravitational wave astronomy. 
\item We need robust and more precise measurements of $\eta$. At present, there 
are several sources with $\eta > 0.32$ at high masses, but we cannot really exclude 
they actually have $\eta < 0.32$~\cite{dl}. There are some sources of uncertainty in the
final estimate of the radiative efficiency and the most important one is the 
determination of the mass of the BH candidate. Moreover, the results presented
in~\cite{dl} are based on a preliminary study and it is necessary to further investigate
and verify the validity of the method proposed by these authors.
\end{enumerate}

\begin{acknowledgments}
This work was supported by the Humboldt Foundation.
\end{acknowledgments}


\appendix

\section{JP metric} \label{a-1}

The JP metric is not a solution in any known gravity theory. It is a simple
parametrization to describe the space-time around non-Kerr BHs and 
was specifically proposed to test the Kerr BH hypothesis~\cite{jp}. 
The metric was obtained by starting from a deformed Schwarzschild 
solution and then by applying a Newman-Janis transformation. The 
non-zero metric coefficients in Boyer-Lindquist coordinates are:
\be
g_{tt} &=& - \left(1 - \frac{2 M r}{\rho^2}\right) (1 + h)
 \, , \nonumber\\
g_{t\phi} &=& - \frac{2 a M r \sin^2\theta}{\rho^2} 
(1 + h) \, , \nonumber\\
g_{\phi\phi} &=& \sin^2\theta \left[r^2 + a^2
+ \frac{2 a^2 M r \sin^2\theta}{\rho^2} \right] + \nonumber\\
&& + \frac{a^2 (\rho^2 + 2 M r) \sin^4\theta}{\rho^2} 
h \, , \nonumber\\
g_{rr} &=& \frac{\rho^2 (1 + h)}{\Delta + 
a^2 h \sin^2\theta } \, , \nonumber\\
g_{\theta\theta} &=& \rho^2 \, ,
\ee
where
\be
\rho^2 &=& r^2 + a^2 \cos^2\theta \, , \nonumber\\
\Delta &=& r^2 - 2 M r + a^2 \, , \nonumber\\
h &=& \sum_{k = 0}^{\infty} \left(\epsilon_{2k} 
+ \frac{M r}{\rho^2} \epsilon_{2k+1} \right)
\left(\frac{M^2}{\rho^2}\right)^k \, .
\ee
The metric has an infinite number of free parameters $\epsilon_i$ and 
the Kerr solution is recovered when all these parameters are set to zero. 
However, in order to recover the correct Newtonian limit we have to 
impose $\epsilon_0 = \epsilon_1 = 0$, while $\epsilon_2$ is constrained 
at the level of $10^{-4}$ from current tests in the Solar System~\cite{jp}.

\section{MN metric} \label{a-2}

The MN metric is a stationary, axisymmetric, and asymptotically flat exact 
solution of the Einstein's vacuum equations with arbitrary mass-multipole 
moments~\cite{mn}. It does not describe the space-time around a BH and
it has naked singularities and closed time-like curves at small radii. These 
pathological features should be either inside some sort of exotic compact 
object, whose exterior gravitational field would be described by the MN 
metric, or GR should break down close to them. The non-zero metric 
coefficients in prolate spheroidal coordinates are:
\be
g_{tt} &=& - f \, , \nonumber\\
g_{t\phi} &=& f \omega \, , \nonumber\\
g_{\phi\phi} &=& \frac{k^2}{f} \left(x^2 - 1\right)
\left(1 - y^2\right) - f \omega^2 \, , \nonumber\\
g_{xx} &=& \frac{k^2 e^{2\gamma}}{f}\frac{x^2 - y^2}{x^2 - 1} \, , \nonumber\\
g_{yy} &=& \frac{k^2 e^{2\gamma}}{f}\frac{x^2 - y^2}{1 - y^2} \, ,
\ee
where
\be
f &=& e^{2\psi} A/B\, , \nonumber\\
\omega &=& 2 k e^{- 2\psi} C A^{-1} 
- 4 k \alpha \left(1 - \alpha^2\right)^{-1} \, , \nonumber\\
e^{2\gamma} &=& e^{2\gamma'}A \left(x^2 - 1\right)^{-1}
\left(1 - \alpha^2\right)^{-2} \, ,
\ee
and
\begin{widetext}
\begin{eqnarray}
\psi &=& \sum_{n = 1}^{+\infty} \frac{\alpha_n P_n}{R^{n+1}} \, , \nonumber\\
\gamma' &=& \frac{1}{2} \ln\frac{x^2 - 1}{x^2 - y^2}  
+ \sum_{m,n = 1}^{+\infty} \frac{(m+1)(n+1) \alpha_m 
\alpha_n}{(m+n+2) R^{m+n+2}} (P_{m+1} P_{n+1} - P_m P_n) 
+ \nonumber\\
&& + \Bigg[ \sum_{n=1}^{+\infty} \alpha_n \Bigg((-1)^{n+1} - 1
+ \sum_{k = 0}^{n} \frac{x-y+(-1)^{n-k}(x+y)}{R^{k+1}}
P_k \Bigg) \Bigg] \, , \nonumber\\
A &=& (x^2 - 1)(1 + ab)^2 - (1 - y^2)(b - a)^2 \, , \nonumber\\
B &=& [x + 1 + (x - 1)ab]^2 + [(1 + y)a + (1 - y)b]^2 \, , \nonumber\\
C &=& (x^2 - 1)(1 + ab)[b - a - y(a + b)]
+ (1 - y^2)(b - a)[1 + ab + x(1 - ab)] \, , \nonumber\\
a &=& -\alpha \exp \left[\sum_{n=1}^{+\infty} 2\alpha_n 
\left(1 - \sum_{k = 0}^{n} \frac{(x - y)}{R^{k+1}} P_k\right)\right] \, , \nonumber\\
b &=& \alpha \exp \Bigg[\sum_{n=1}^{+\infty} 2\alpha_n 
\Bigg((-1)^n + \sum_{k = 0}^{n} \frac{(-1)^{n-k+1}(x + y)}{R^{k+1}} P_k\Bigg)\Bigg] \, .
\end{eqnarray}
\end{widetext}
Here $R = \sqrt{x^2 + y^2 - 1}$ and $P_n$ are the Legendre 
polynomials with argument $xy/R$,
\be
P_n &=& P_n\left(\frac{xy}{R}\right) \, , \nonumber\\
P_n(\chi) &=& \frac{1}{2^n n!} \frac{d^n}{d\chi^n} 
\left(\chi^2 - 1\right)^n \, .
\ee
The standard Boyer-Lindquist coordinates $(r,\theta)$ are related to 
the prolate spheroidal coordinates $(x,y)$ by
\be
r &=& k x + M \, , \nonumber\\ 
\cos\theta &=& y \, .
\ee

The MN solution has an infinite number of free parameters: $k$, which 
regulates the mass of the space-time; $\alpha$, which regulates the 
spin; and $\alpha_n$ ($n=1, . . . , +\infty$) which regulates the mass-multipole 
moments, starting from the dipole $\alpha_1$, to the quadrupole  
$\alpha_2$, etc. For $\alpha \neq 0$ and $\alpha_n = 0$, the MN solution 
reduces to the Kerr metric. For $\alpha = \alpha_n = 0$, it reduces to the 
Schwarzschild solution. For $\alpha = 0$ and $\alpha_n \neq 0$, one 
obtains the static Weyl metric. Without loss of generality, we can put 
$\alpha_1 = 0$ to bring the massive object to the origin of the coordinate 
system. The simplest extension of the Kerr solution is thus the subclass of 
MN space-times with $\alpha_n = 0$ for $n \neq 2$. Here there are three 
free parameters ($k$, $\alpha$, and $\alpha_2$) related to the mass $M$, 
the dimensionless spin parameter $a_* = J/M^2$, and the dimensionless 
anomalous quadrupole moment $\tilde{q}$, defined by $Q = - (1+\tilde{q})
a_*^2M^3$, with $Q$ the mass-quadrupole moment of the object, by the 
relations
\be
\alpha &=& \frac{\sqrt{1 - a^2_*} - 1}{a_*} \, , \nonumber\\
k &=& M \frac{1 - \alpha^2}{1 + \alpha^2} \, , \nonumber\\
\alpha_2 &=& \tilde{q} a_*^2 \frac{M^3}{k^3} \, .
\ee
Note that $\tilde{q}$ measures the deviation from the quadrupole
moment of a Kerr BH. In particular, since $Q_{\rm Kerr} = - a^2_* M^3$, 
the solution is oblate (prolate) for $\tilde{q} > - 1$ ($\tilde{q} < - 1$), but 
it is more oblate (prolate) than the Kerr one for $\tilde{q} > 0$ ($\tilde{q} < 0$). 
When $\tilde{q}=0$, the solution reduces to the Kerr metric, but when 
$\tilde{q} \neq 0$ also the higher-order mass-multipole moments have a 
different value than in Kerr.




\begin{thebibliography}{99}

\bibitem{w}
  C.~M.~Will,
  Living Rev.\ Rel.\  {\bf 9}, 3 (2006)
  [gr-qc/0510072].
  
\bibitem{5th} 
  E.~Fischbach and C.~L.~Talmadge,
  {\it The search for nonNewtonian gravity},
  (Springer, New York, New York, 1999).
  
\bibitem{p}
  R.~Penrose,
  Riv.\ Nuovo Cim.\  {\bf 1}, 252 (1969)
  [Gen.\ Rel.\ Grav.\  {\bf 34}, 1141 (2002)].
  
\bibitem{c} 
  B.~Carter,
  Phys.\ Rev.\ Lett.\  {\bf 26}, 331 (1971);
  D.~C.~Robinson,
  Phys.\ Rev.\ Lett.\  {\bf 34}, 905 (1975).
  
\bibitem{n}
  R.~Narayan,
  New J.\ Phys.\  {\bf 7}, 199 (2005)
  [gr-qc/0506078].
  
\bibitem{kb}
  V.~Kalogera and G.~Baym,
  Astrophys.\ J.\  {\bf 470}, L61 (1996)
  [astro-ph/9608059].
  
\bibitem{m}
  E.~Maoz,
  Astrophys.\ J.\  {\bf 494}, L181 (1998)
  [astro-ph/9710309].
  
\bibitem{bln} 
  R.~Narayan and J.~E.~McClintock,
  New Astron.\ Rev.\  {\bf 51}, 733 (2008)
  [arXiv:0803.0322 [astro-ph]];
  A.~E.~Broderick, A.~Loeb and R.~Narayan,
  Astrophys.\ J.\  {\bf 701}, 1357 (2009)
  [arXiv:0903.1105 [astro-ph.HE]].
  
\bibitem{a}
  M.~A.~Abramowicz, W.~Kluzniak and J.~-P.~Lasota,
  Astron.\ Astrophys.\  {\bf 396}, L31 (2002)
  [astro-ph/0207270].
  
\bibitem{b4}
  C.~Bambi,
  Mod.\ Phys.\ Lett.\ A {\bf 26}, 2453 (2011)
  [arXiv:1109.4256 [gr-qc]].
 
\bibitem{r}
  F.~D.~Ryan,
  Phys.\ Rev.\ D {\bf 52}, 5707 (1995);
  K.~Glampedakis and S.~Babak,
  Class.\ Quant.\ Grav.\  {\bf 23}, 4167 (2006)
  [gr-qc/0510057]. 
  L.~Barack and C.~Cutler,
  Phys.\ Rev.\ D {\bf 75}, 042003 (2007)
  [gr-qc/0612029];
  T.~A.~Apostolatos, G.~Lukes-Gerakopoulos and G.~Contopoulos,
  Phys.\ Rev.\ Lett.\  {\bf 103}, 111101 (2009)
  [arXiv:0906.0093 [gr-qc]]. 
  
\bibitem{bb1}
  C.~Bambi and E.~Barausse,
  Astrophys.\ J.\  {\bf 731}, 121 (2011)
  [arXiv:1012.2007 [gr-qc]].
  
\bibitem{pj}
  D.~Psaltis and T.~Johannsen,
  arXiv:1011.4078 [astro-ph.HE].  
  
\bibitem{jpp}
  T.~Johannsen and D.~Psaltis,
  Astrophys.\ J.\  {\bf 726}, 11 (2011)
  [arXiv:1010.1000 [astro-ph.HE]].
  
\bibitem{b2}
  C.~Bambi,
  Phys.\ Rev.\ D {\bf 83}, 103003 (2011)
  [arXiv:1102.0616 [gr-qc]].
  
\bibitem{b5}
  C.~Bambi,
  Phys.\ Lett.\ B {\bf 705}, 5 (2011)
  [arXiv:1110.0687 [gr-qc]].
  
\bibitem{bf}
  C.~Bambi and K.~Freese,
  Phys.\ Rev.\ D {\bf 79}, 043002 (2009)
  [arXiv:0812.1328 [astro-ph]];
  C.~Bambi and N.~Yoshida,
  Class.\ Quant.\ Grav.\  {\bf 27}, 205006 (2010)
  [arXiv:1004.3149 [gr-qc]];
  C.~Bambi, F.~Caravelli and L.~Modesto,
  arXiv:1110.2768 [gr-qc].
  
\bibitem{bb2}
  C.~Bambi and E.~Barausse,
  Phys.\ Rev.\ D {\bf 84}, 084034 (2011)
  [arXiv:1108.4740 [gr-qc]].

\bibitem{bm}
  C.~Bambi and L.~Modesto,
  Phys.\ Lett.\ B {\bf 706}, 13 (2011)
  [arXiv:1107.4337 [gr-qc]].

\bibitem{s} 
  A.~Soltan,
  Mon.\ Not.\ Roy.\ Astron.\ Soc.\  {\bf 200}, 115 (1982).  
  
\bibitem{erz}
  M.~Elvis, G.~Risaliti and G.~Zamorani,
  Astrophys.\ J.\  {\bf 565}, L75 (2002)
  [astro-ph/0112413].
  
\bibitem{ho}
  J.~-M.~Wang, Y.~-M.~Chen, L.~C.~Ho and R.~J.~McLure,
  Astrophys.\ J.\  {\bf 642}, L111 (2006)
  [astro-ph/0603813]. 
  
\bibitem{dl} 
  S.~W.~Davis and A.~Laor,
  Astrophys.\ J.\  {\bf 728}, 98 (2011)
  [arXiv:1012.3213 [astro-ph.CO]].

\bibitem{jp}
  T.~Johannsen and D.~Psaltis,
  Phys.\ Rev.\ D {\bf 83}, 124015 (2011)
  [arXiv:1105.3191 [gr-qc]].   
  
\bibitem{mn}
  V.~S.~Manko and I.~D.~Novikov,
  Class.\ Quant.\ Grav.\  {\bf 9}, 2477 (1992).
  
\bibitem{nt}
  I.~D.~Novikov, K.~S.~Thorne,
  ``Astrophysics of Black Holes'' in {\it Black Holes}, 
  edited by C.~De~Witt and B.~De~Witt
  (Gordon and Breach, New York, New York, 1973), pp. 343-450;
  D.~N.~Page and K.~S.~Thorne,
  Astrophys.\ J.\  {\bf 191}, 499 (1974).
  
\bibitem{ss}
  N.~I.~Shakura and R.~A.~Sunyaev,
  Astron.\ Astrophys.\  {\bf 24}, 337 (1973).
  
\bibitem{bp} 
  J.~M.~Bardeen and J.~A.~Petterson,
  Astrophys.\ J.\  {\bf 195}, L65 (1975).

\bibitem{b1} 
  C.~Bambi,
  Europhys.\ Lett.\  {\bf 94}, 50002 (2011)
  [arXiv:1101.1364 [gr-qc]].
  
\bibitem{t} 
  K.~S.~Thorne,
  Astrophys.\ J.\  {\bf 191}, 507 (1974).

\bibitem{li}
  L.~-X.~Li, E.~R.~Zimmerman, R.~Narayan and J.~E.~McClintock,
  Astrophys.\ J.\ Suppl.\  {\bf 157}, 335 (2005)
  [astro-ph/0411583].
  
\bibitem{ak}
  E.~Agol and J.~Krolik,
  Astrophys.\ J.\  {\bf 528}, 161 (2000)
  [astro-ph/9908049];
  A.~Tchekhovskoy, R.~Narayan and J.~C.~McKinney,
  arXiv:1108.0412 [astro-ph.HE].
  
\bibitem{g} 
  C.~F.~Gammie, S.~L.~Shapiro and J.~C.~McKinney,
  Astrophys.\ J.\  {\bf 602}, 312 (2004)
  [astro-ph/0310886].
  
\bibitem{k}
  J.~H.~Krolik,
  Astrophys.\ J.\  {\bf 515}, L73 (1999)
  [astro-ph/9902267].
  
\bibitem{n09}
  S.~C.~Noble and J.~H.~Krolik,
  Astrophys.\ J.\  {\bf 703}, 964 (2009)
  [arXiv:0907.1655 [astro-ph.HE]];
  S.~C.~Noble, J.~H.~Krolik and J.~F.~Hawley,
  Astrophys.\ J.\  {\bf 711}, 959 (2010)
  [arXiv:1001.4809 [astro-ph.HE]].
  
\bibitem{ap}
  N.~Afshordi and B.~Paczynski,
  Astrophys.\ J.\  {\bf 592}, 354 (2003)
  [astro-ph/0202409].
  
\bibitem{mcc}
  J.~E.~McClintock, R.~Shafee, R.~Narayan, R.~A.~Remillard, S.~W.~Davis and L.~-X.~Li,
  Astrophys.\ J.\  {\bf 652}, 518 (2006)
  [astro-ph/0606076].

\bibitem{p10} 
  R.~F.~Penna, J.~C.~McKinney, R.~Narayan, A.~Tchekhovskoy, R.~Shafee and J.~E.~McClintock,
  Mon.\ Not.\ Roy.\ Astron.\ Soc.\  {\bf 408}, 752 (2010)
  [arXiv:1003.0966 [astro-ph.HE]].
  
\bibitem{p11}
  R.~F.~Penna, A.~Sadowski and J.~C.~McKinney,
  arXiv:1110.6556 [astro-ph.HE].
  
\bibitem{th}
  R.~Takahashi and T.~Harada,
  Class.\ Quant.\ Grav.\  {\bf 27}, 075003 (2010)
  [arXiv:1002.0421 [astro-ph.HE]].
  
\bibitem{b70} 
  J.~M.~Bardeen,
  Nature {\bf 226}, 64 (1970).
  
\bibitem{b3}
  C.~Bambi,
  JCAP {\bf 1105}, 009 (2011)
  [arXiv:1103.5135 [gr-qc]].

\bibitem{glm}
  J.~R.~Gair, C.~Li and I.~Mandel,
  Phys.\ Rev.\ D {\bf 77}, 024035 (2008)
  [arXiv:0708.0628 [gr-qc]].

\bibitem{mms1}
  V.~S.~Manko, E.~W.~Mielke and J.~D.~Sanabria-Gomez,
  Phys.\ Rev.\ D {\bf 61}, 081501 (2000)
  [gr-qc/0001081];
  V.~S.~Manko, J.~D.~Sanabria-Gomez and O.~V.~Manko,
  Phys.\ Rev.\ D {\bf 62}, 044048 (2000).

\bibitem{lp}
  W.~G.~Laarakkers and E.~Poisson,
  Astrophys.\ J.\  {\bf 512}, 282 (1999)
  [gr-qc/9709033]. 

\end{thebibliography}
\end{document}